\documentclass[12pt]{iopart}
\expandafter\let\csname equation*\endcsname\relax
\expandafter\let\csname endequation*\endcsname\relax
\usepackage{amsfonts}
\usepackage{bbold}
\usepackage{amsmath}
\usepackage{iopams}
\usepackage{textcomp}
\usepackage{amssymb}
\usepackage{mathtools}
\usepackage{cancel} 
\usepackage{bbm}
\usepackage[usenames, dvipsnames]{color} 
\usepackage{dcolumn}
\usepackage{cancel, soul,ulem}
\usepackage{epic} 
\usepackage{epsfig}
\usepackage{feynmp}
\usepackage{graphicx}
\usepackage{cite}
\usepackage{grffile}
\usepackage[breaklinks,colorlinks = true,linkcolor = red,urlcolor=blue,citecolor=cyan]{hyperref}
\usepackage{mathrsfs}
\usepackage[mathscr]{eucal}
\usepackage{soul}
\usepackage{subfigure}
\usepackage{wrapfig}
\usepackage{xy} 
\usepackage{xcolor}
\usepackage{amsmath,amssymb,amsfonts,amsthm}
\usepackage[ascii]{inputenc}
\usepackage{makecell}
\usepackage{titlesec}

\makeatletter
\renewcommand\@appendixstar{\@@par
 \ifnumbysec 
 \@addtoreset{table}{section}
 \@addtoreset{figure}{section}\fi
 \setcounter{section}{0}
 \setcounter{subsection}{0}
 \setcounter{subsubsection}{0}
 \setcounter{equation}{0}
 \setcounter{figure}{0}
 \setcounter{table}{0}
 \def\thesection{\Alph{section}}
 \def\theequation{\ifnumbysec
      \Alph{section}.\arabic{equation}\else
      \Alph{section}\arabic{equation}\fi}
 \def\thetable{\ifnumbysec
      \Alph{section}\arabic{table}\else
      A\arabic{table}\fi}
 \def\thefigure{\ifnumbysec
      \Alph{section}\arabic{figure}\else
      A\arabic{figure}\fi}}
\makeatother

\begin{document}

\title{Full counting statistics of 1d short-range Riesz gases in confinement}

\author{Jitendra Kethepalli}
\address{International Centre for Theoretical Sciences, Tata Institute of Fundamental Research, Bengaluru -- 560089, India}
\author{Manas Kulkarni}
\address{International Centre for Theoretical Sciences, Tata Institute of Fundamental Research, Bengaluru -- 560089, India}
\author{Anupam Kundu}
\address{International Centre for Theoretical Sciences, Tata Institute of Fundamental Research, Bengaluru -- 560089, India}
\author{Satya N. Majumdar}
\address{LPTMS, CNRS, Univ.  Paris-Sud,  Universit{\'e} Paris-Saclay,  91405 Orsay,  France}
\author{David Mukamel}
\address{Department of Physics of Complex Systems, Weizmann Institute of Science, Rehovot 7610001, Israel}
\author{Gr\'egory Schehr}
\address{Sorbonne Universit{\'e}, Laboratoire de Physique Th{\'e}orique et Hautes Energies, CNRS UMR 7589, 4 Place Jussieu, 75252 Paris Cedex 05, France}

\newpage
\begin{abstract}
We investigate the full counting statistics (FCS) of a harmonically confined 1d short-range Riesz gas consisting of $N$ particles in equilibrium at finite temperature. The particles interact with each other through a repulsive power-law interaction with an exponent $k>1$ which includes the Calogero-Moser model for $k=2$. We examine the probability distribution of the number of particles in a finite domain $[-W, W]$ called number distribution, denoted by $\mathcal{N}(W, N)$. We analyze the probability distribution of $\mathcal{N}(W, N)$ and show that it exhibits a large deviation form for large $N$ characterised by a speed $N^{\frac{3k+2}{k+2}}$ and by a large deviation function of the fraction $c = \mathcal{N}(W, N)/N$ of the particles inside the domain and $W$. We show that the density profiles that create the large deviations display interesting shape transitions as one varies $c$ and $W$. This is manifested by a third-order phase transition exhibited by the large deviation function that has discontinuous third derivatives. Monte-Carlo (MC) simulations show good agreement with our analytical expressions for the corresponding density profiles. We find that the typical fluctuations of $\mathcal{N}(W, N)$, obtained from our field theoretic calculations are Gaussian distributed with a variance that scales as $N^{\nu_k}$, with $\nu_k = (2-k)/(2+k)$. We also present some numerical findings on the mean and the variance. Furthermore, we adapt our formalism to study the index distribution (where the domain is semi-infinite $(-\infty, W])$, linear statistics (the variance), thermodynamic pressure and bulk modulus.

\end{abstract}

\date{\today}

\maketitle

\section{Introduction}
\label{sec1}
The study of many-particle low-dimensional quantum and classical systems have been a subject of great theoretical and experimental interest. A very interesting observable that unravels the equilibrium and non-equilibrium properties of low-dimensional systems is the distribution of the number of particles in a given domain. This is often referred to as the full counting statistics (FCS). Since FCS, the total number of particles in a domain, is a global quantity it is experimentally more accessible ~\cite{esteve2006observations,jacqmin2011sub}.

In the context of quantum systems, FCS has been studied in various physical setups, including non-equilibrium Luttinger liquids~\cite{protopopov2012luttinger}, quantum transport~\cite{levitov1996electron, Beenaker1993B, kanzieper1}, shot noise~\cite{levitov1993charge, vivo2008distribution, kanzieper2, sommers, bohigas1}, quantum dots~\cite{groth2006counting, Grabsch3}  as well as in quantum spin chains and fermionic chains~\cite{eisler2013full,ivanov2013characterizing}. Furthermore, the entanglement entropy of a subsystem with its remaining part, studied extensively in the context of the free Fermi gas, has been found to be intricately connected to FCS~\cite{klich2006lower, klich2009quantum, song2011entanglement, calabrese2015random, majumdar2018entanglement, smith2021counting}. This connection holds true, particularly in regimes where the particle number fluctuations exhibit Gaussian behaviour.  The study of FCS for interacting systems has also gained considerable interest~\cite{bastianello2018sinh, arzamasovs2019full, smith2021full, bastianello2018exact} as FCS can now be measured in cold atom experiments ~\cite{esteve2006observations,jacqmin2011sub}. This connection emphasizes the wide-ranging applications of FCS, particularly in understanding the relationship between interactions and correlations in the system.

In the context of classical systems, the  FCS has also been widely investigated. For instance, in many ecological settings, it has been observed that the distribution of the number of species and the average number of species in a given domain exhibit universal features~\cite{law2009ecological, azaele2016statistical, akemann2023interactions}. The statistics of the number of particles in specific domains for different point processes have also been investigated~\cite{shirai2003random,ghosh2018point}. Such point processes can be classified based on the system size dependence of the Fano factor $\mathbb{V}(\mathcal{D}) = {\rm Var}\left(\mathcal{N}\left({\mathcal{D}}\right)\right)/\langle \mathcal{N}\left({\mathcal{D}}\right)\rangle$ where $\mathcal{N}\left({\mathcal{D}}\right)$ is the number of particles in a given domain $\mathcal{D}$. This ratio of the variance and the mean 
measures the strength of the relative fluctuations of $\mathcal{N}\left({\mathcal{D}}\right)$. In the large $N$ limit, typical systems such as Poissonian point processes are characterized by $\mathbb{V}(\mathcal{D}) \sim O(1)$. Interestingly there are some systems for which $\mathbb{V}(\mathcal{D}) \to 0$, in the large $N$ limit and they are generically called hyperuniform~\cite{torquato2016hyperuniformity, torquato2018hyperuniform}. 

While FCS is an interesting quantity both in classical and quantum systems the role of interactions are not well understood. This article investigates FCS in a one-dimensional system of classical particles interacting via a power-law potential known as the  Riesz gas~\cite{riesz,lewin}. We consider a harmonically confined Riesz gas composed of $N$ particles in thermal equilibrium described by the Boltzmann distribution $P(\{x_i\}) = \exp[-\beta \tilde{E}(\{x_i\})]/Z_N$ where $\beta^{-1}$ is temperature and $Z_N$ is the partition function. The energy function of the gas is given by
\begin{equation}\label{hamiltonian}
     \tilde{E}_k(\{x_i\}) = \sum_i^N \frac{x_i^2}{2} + \frac{J \, \text{sgn}(k)}{2} \sum_{i=1}^N\sum_{j \neq i}^N |x_i-x_j|^{-k}, 
\end{equation}
where $x_i$ is the position of $i^{\rm th}$ particle with $i=1,2,\cdots,N$ and $\text{sgn}(k)$ is the sign function. The strength of the repulsive interaction is controlled by $J>0$ and the exponent $k$ of the power law determines the nature of interactions; in particular, for $k>1$ the system is short-ranged and for $k<1$ it is long-ranged. Interaction with certain values of $k$ correspond to well known models. For instance, $k=2$ corresponds to the Calogero-Moser model ~\cite{calogero1971solution, calogero1975exactly, sutherland, moser, agarwalCM} which represents an integrable many-particle interacting system.  The gas of hard rods corresponds to $k \to \infty$~\cite{percus1976equlibrium, kethepalli2022finite}, Dyson's log-gas has $k \to 0$ and $J \to J_0/k$~\cite{Dyson1962,Dyson1963,Mehtabook,wigner1958distribution} and the 1d one component plasma (1dOCP) has $k=-1$~\cite{dhar2018extreme, Flack2022, flack2022exact, lenard61, Baxter1963, dhar2017exact, Flack2021,  Chafai22, flack2023out, rojas2018universal}. Additionally, some fractional values of $k$ have also been experimentally realized~\cite{joseph2011observation, zhang2017observation}. The statistical properties of the Riesz gas model have gained considerable interest in recent years~\cite{leble2017, leble2018, agarwal2019harmonically, lewin, dereudre2020existence, kumar2020particles, boursier2021optimal, jit2021, boursier2022decay, santra2022, jit2022, lelotte2023phase, santra2023crossover, dereudre2023number, dandekar2023dynamical}. In this paper we study FCS of the Riesz gas defined in Eq.~\eqref{hamiltonian} and we restrict ourselves to the short-range interactions, {\it i.e.}, $k>1$ where the associated field theory is local~\cite{agarwal2019harmonically}. It is to be noted that for the Riesz gas, the exact results on FCS are only known for $k \to 0$~\cite{dyson1963statistical, FS, bai2009clt, borodin14,marino14, marino2016number} and $k=-1$~\cite{dhar2018extreme,Flack2022}, both of which are long-range models.

The rest of the paper is organized as follows. In Section~\ref{sec2}, we introduce the relevant quantities and notations that we use throughout the paper. In Section~\ref{sec3}, we summarize our main results on FCS of the Riesz gas. In Section~\ref{sec4}, we explain the derivation of the large deviation function, which characterizes the probability distribution of the number of particles in the domain $[-W, W]$. The corresponding average density profiles are also calculated. These profiles are used to study the variance along with asymptotic behaviours, and non-analytic properties of the associated large deviation function (LDF). Our formalism has been adapted to study the index distribution which corresponds to the semi-infinite domain $(-\infty, W]$ in Section~\ref{indist}. In Section~\ref{sec5}, we study the linear statistics of the Riesz gas. We conclude and provide some future directions in Section~\ref{sec6}.

\begin{figure}
    \centering
    \includegraphics{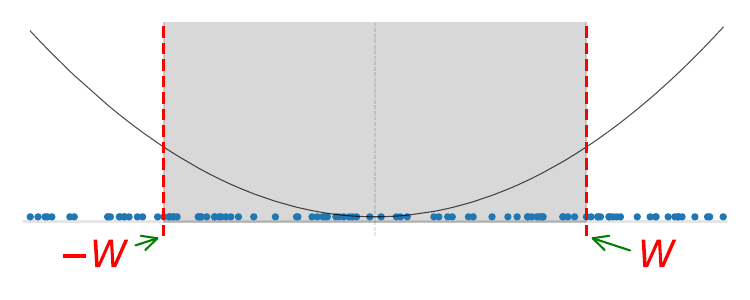}
    \caption{Schematic representation of the domain $[-W, W]$ (shaded region) studied in this paper. The blue dots are the positions of the particles. The number of particles in the region $[-W, W]$ is defined as $\mathcal{N}(W, N)$. The black solid line is indicative of the harmonic confinement.}
    \label{fig:schembox}
\end{figure}

\section{Relevant properties of Riesz gas and notations}
\label{sec2}
In this section, we recap some results and properties of short-range Riesz gas, with $k>1$ which were first derived in Ref.~\cite{agarwal2019harmonically}. These results form the foundation of our present work. Since $\beta \sim O(1)$, entropy does not play a dominant role in competing with the confining potential; instead, the repulsive interaction plays the dominant role. As a result, the particles settle down over a finite support of length $L_N$ which scales for large-$N$ as ~\cite{agarwal2019harmonically, jit2021}
\begin{equation}
L_N \sim O\left(N^{\alpha_k}\right) \quad \text{with} \quad \alpha_k = \frac{k}{k+2}  \text{ for } k>1 \, . \label{L_N-alpha_k}
\end{equation}
It is convenient to scale the positions of the particles by this length scale $N^{\alpha_k}$ i.e. 
\begin{align}\label{scalexy}
y_i = \frac{x_i}{N^{\alpha_k}},
\end{align}
with $y_i \sim O(1)$. 
In terms of these new scaled variables, the energy function given in Eq.~\eqref{hamiltonian} now reads 
\begin{align}\label{shamiltonian}
     \tilde{E}_k(\{y_i N^{\alpha_k}\}) &=N^{1+2\alpha_k} E_k(\{y_i\}),\notag ~~~~~~\text{where,}~\\
     E_k(\{y_i\}) &= \left(\frac{1}{N}\sum_i^N \frac{y_i^2}{2} + \frac{1}{N^{1+k}}\frac{J \, \text{sgn}(k)}{2} \sum_{i=1}^N\sum_{j \neq i}^N |y_i-y_j|^{-k}\right). 
\end{align}
Since, $y_i \sim O(1)$,  the rescaled energy $E_k(\{y_i\})$ is also of order $O(1)$. 
As described in the Ref.~\cite{agarwal2019harmonically}, one can express the energy function  in Eq.~\eqref{shamiltonian} as a functional of the density profile $\rho(y) = \tfrac{1}{N}\sum_{i=1}^N \delta(y-y_i)$. For the short-range Riesz gas, the energy functional takes the following form \cite{agarwal2019harmonically}
\begin{align}
\mathcal{E}_k\left[\rho(y)\right] \approx \int_{-\infty}^{\infty} dy~ \frac{y^{2}}{2}\rho(y) +J \zeta(k)  \int_{-\infty}^{\infty} dy~\rho(y)^{k+1} + O(1/N),~~\text{for}~k>1.
\label{E-func-k>1}
\end{align}
This field theory was subsequently used to compute the average density profile via a saddle point approximation. The average density profile in terms of the scaled variables $y$  is explicitly given by~\cite{agarwal2019harmonically, hardin2018large}
\begin{equation} \label{rho_uc}
    \rho_{0}(y) \equiv \rho_{k, \rm uc}^*(y) = A_k\left(l_0^2-y^2\right)^{\frac{1}{k}},~~\text{for}~ |y|\leq l_0,
\end{equation}
with 
\begin{equation}
l_0 \equiv l_k^{\rm uc} = \Bigg(A_k~ B\left(\frac{1}{2}, \frac{1}{k}+1\right)\Bigg)^{-\alpha_k} 
~ \text{along with} ~
A_k = \left(2J(k+1) \zeta(k)\right)^{-\frac{1}{k}},~\forall ~k>1.
\label{A_1k}
\end{equation}
In Eq.~\eqref{A_1k}, $l_k^{\rm uc}$ represents the edge of the support of the unconstrained density profile $\rho_{k, \rm uc}^*(y)$ {\it i.e.}, in the absence of any additional barriers. Note that the notations with subscript/superscript was originally used in Ref.~\cite{jit2021}, where the superscript ``uc" stood for ``unconstrained". Here $B(x,y)$ is the Beta function and $\zeta(k) = \sum_{n=1}^{\infty} 1/n^{k}$ is the Riemann Zeta function. 

Recall that, our aim in this paper is to study the statistical properties of $\mathcal{N}(W, N)$ which represents the number of particles in domain $[-W,  W]$. As will be discussed later, in the large-$N$ limit, the problem of finding the distribution of $\mathcal{N}(W, N)$ at $O(1)$ temperature gets effectively converted to an optimization problem. This problem tries to find the most probable density profile satisfying the constraint of fixed $\mathcal{N}(W, N)$. Note that, under the transformation to the rescaled variables in Eq.~\eqref{scalexy}, the wall position $W$ gets transformed to $w = W/N^{\alpha_k}$. It is evident that if $w > l_0$ then the density profile does not get affected by the presence of the hard walls and it remains the unconstrained density profile given in Eq.~\eqref{rho_uc}. On the other hand for $w<l_0$ the most probable density profile will be drastically different from the one given in Eq.~\eqref{rho_uc}.  This modified density profile, as we will see later, is an important ingredient for the study of FCS. We compute the constrained density profile and use it to study the probability distribution (more precisely the associated LDF) of $\mathcal{N}(W,N)$. Before going into the details of the computation, we summarize our main findings in the next section.

\section{Summary of the main results}
\label{sec3}
\begin{figure}
    \centering
    \includegraphics[scale=0.6]{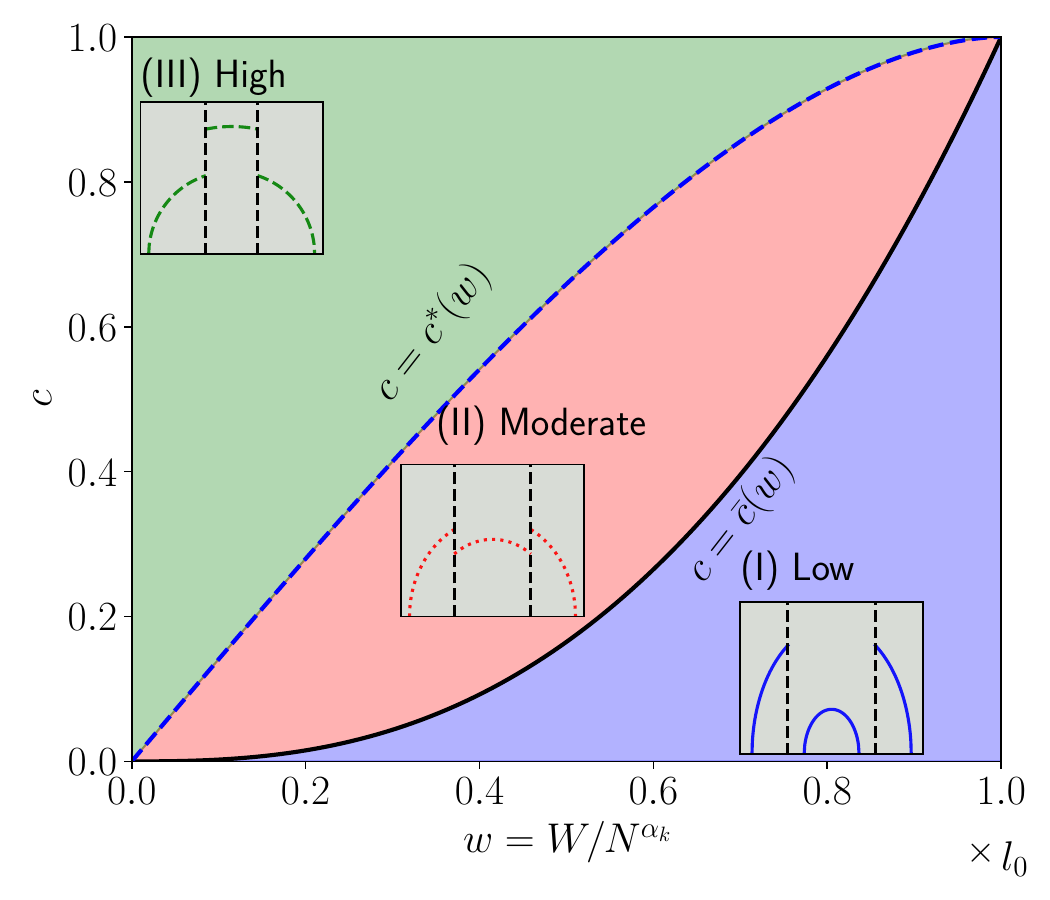}
    \caption{Phase diagram in the $(w, c)$ plane showing three different regimes: (I) low, (II) moderate and (III) high fraction of particles in the box $[-W, W]$, where the saddle point density profiles given in Eq.~\eqref{derv:rho} exhibit distinctly different shapes (see inset). The critical fraction line $c = \bar{c}(w)$ [Eq.~\eqref{cbar}] separates the phases (I) and (II). Below this fraction, we observe a disjoint density profile in phase (I) and above this concentration the two disjoint parts join and we observe a density profile as shown in the inset of regime (II).  Around this line, the LDF behaves non-analytically which leads to a third-order phase transition [see Appendix.~\ref{append:asymptotic}]. On the other hand, the crossover line $c = c^*(w)$ [Eq.~\eqref{cstarn0}] separates the phases (II) and (III) and the LDF shows analytic behaviour around it [see Eq.~\eqref{derv:number-prob-typical}]. This plot is generated for $k=1.25$, however, such a plot is expected to be qualitatively the same for all $k>1$. Note that the $x$-axis is in the units of $l_0$.}
    \label{fig:phase}
\end{figure}

In this section, we present the main results related to the statistics of $\mathcal{N}(W,N)$, the number of particles in a finite box $[-W, W]$ which is schematically shown in Fig.~\ref{fig:schembox}. It is easy to show that the mean number of particles in the box increases linearly with the system size (i.e., number of particles $N$) and is given by
\begin{align}\label{cstarn0}
\langle \mathcal{N}(W, N) \rangle  &\simeq N~c^*\left(\frac{W}{N^{\alpha_k}}\right)\\
~\text{with}~&~c^*(w)= \int_{-w}^w dy~\rho_{0}(y), \label{def:c^*}
\end{align}
where the unconstrained density profile $\rho_{0}(y)$ is given in Eq.~\eqref{rho_uc}. We denote the probability distribution of $\mathcal{N}(W,N)$ as
\begin{align}
    \mathscr{P}\left(\mathcal{N} = c\,N, W \right) = {\rm Prob}.[\mathcal{N}(W, N)=c\,N,W].
\end{align}
We find that in the large-$N$ limit, the probability distribution takes the large deviation form given by 
\begin{align}
	\mathscr{P}\left(\mathcal{N} = c~N, W\right) &\asymp  \exp\Big(-\beta N^{1+2\alpha_k} \Phi(c, W/N^{\alpha_k})\Big), \label{mcal(P)_n}
\end{align}
valid when $W\to\infty$, $N\to\infty$ keeping the ratio $w = W/N^{\alpha_k}$ fixed. Here we recall that $\alpha_k= k/(k+2)$ as given in Eq.~\eqref{L_N-alpha_k}. To calculate the LDF $\Phi(c, W/N^{\alpha_k})$, we use the Coulomb gas method~\cite{dean06, dean08}. A crucial ingredient in this method is the saddle point density profiles that satisfy the constraint of having $c\,N$ particles in the box $[-W, W]$. We find that these constrained density profiles are also dome-shaped similar to the unconstrained density profiles. However, their support is parameterized by the box size controlled by $w=W/N^{\alpha_k}$ and the fraction of particles $c$ inside it. As $w$ and $c$ are varied, the shape of the constrained density profile undergoes interesting shape transitions in the $(w-c)$ plane, as indicated in Fig.~\ref{fig:phase} by shaded regions separated by the two curves $c = \bar{c}(w)$ and $c = c^*(w)$. The loci of these two curves are calculated analytically in  Eq.~\eqref{cbar} and Eq~\eqref{def:c^*} respectively. In Fig.~\ref{fig:densnumber} we compare our analytical results for the saddle point density with the MC simulations and observe a very good agreement.

\noindent
\textit{The LDF and its properties:} Using these saddle point densities, we have obtained explicit analytical expression of the LDF $\Phi(c, w)$ which is given in Eq.~\eqref{derv:number-ldf1} and plotted in Fig.~\ref{fig:phinumber}. The behaviour of the LDF is similarly governed by two parameters $c$ and $w=W/N^{\alpha_k}$. We note that, for a fixed box $[-W, W]$, as the fraction of particles is increased from below $\bar{c}(w)$ [regime (I)] to above it [regime (II)], the hole region in the density profile vanishes (see Fig.~\ref{fig:phase}). This gap closing transition at $c=\bar{c}(w)$ with a fixed $w$ gives rise to a non-analytic behaviour of the LDF characterized by a discontinuous third-order derivative of the LDF $\Phi(c, w)$ w.r.t. $c$ for $|c-\bar{c}(w)| \ll 1$ as can be seen from
\begin{align}\label{ldf3dphivsc}
    \frac{\partial^3\Phi(c,w)}{\partial c^3}&= 
    \begin{cases}
       \frac{(1+2k)(1+k)}{\bar{c}(w)^3k^3}\tilde{\mathbb{C}}_{+}\left(\frac{c-\bar{c}(w)}{\bar{c}(w)}\right)^{\frac{1}{k}-1},&~{\rm for}~c >\bar{c}(w)\\
       \frac{6}{\bar{c}(w)^3}\tilde{\mathbb{C}}_{-},&~{\rm for}~c <\bar{c}(w)
    \end{cases},
\end{align}
where $\tilde{\mathbb{C}}_\pm$ are constants [see Appendix~\ref{append:asymptotic} for details]. This discontinuity in the third derivative implies a third-order phase transition according to Ehrenfest classification ~\cite{hilfer2000applications}. By the same mechanism, a similar gap-closing transition occurs but now with decreasing the box size for a fixed $c$ (along an horizontal line in Fig.~\ref{fig:phase}). Such third-order phase transitions via gap-closing mechanisms has been found in numerous examples ~\cite{majumdar2014top}. 

Similar non-analytic behaviour of the LDF associated to $\mathcal{N}(W, N)$ has also been observed in long-range interacting models such as the Dyson's log-gas \cite{marino14, marino2016number} and the 1dOCP \cite{Flack2022}. Interestingly, the non-analyticity of LDF in our short-range case ($k>1$) of Riesz gas, appears at $c=\bar{c}(w)$ unlike these long-range models ($k\to 0$ and $k=-1$) for which it appears at $c=c^*(w)$. For our short-range case, the LDF $\Phi(c,w)$ is analytic at $c=c^*(w)$ and shows quadratic behavior, i.e., $\Phi(c=c^*(w)+\kappa, w) \sim O(\kappa^2)$. This quadratic behaviour of the LDF $\Phi(c,w)$ around $c=c^*(w)$ suggests that the typical fluctuations in the number of particles in the box are described by a Gaussian probability distribution given by
\begin{align}
    \mathscr{P}\left(\mathcal{N} = c\, N, W \right)& \asymp  \exp\Bigg( \frac{-N^2(c-c^*(w))^2}{2\,{\rm Var}(\mathcal{N})}\Bigg), \\
    & \quad \quad  ~\text{for}~|c-c^*(w)| \lesssim O\left(\sqrt{{\rm Var}(\mathcal{N})}\right). \notag
\end{align}
Here the variance is given by
\begin{align}
    {\rm Var}(\mathcal{N}) &= 
 \frac{N^{\nu_k}}{\beta~l_0^{2}~\alpha_k}~\mathcal{V}\left(\frac{W}{N^{\alpha_k}l_0} \right)~\text{with}~\nu_k = \frac{2-k}{2+k}\label{derv:number-var},
\end{align}
and the function $\mathcal{V}\left(h\right)$ is given in Eq.~\eqref{mcal(N)_1}. The analytical result in Eq.~\eqref{cstarn0} for the mean and in Eq.~\eqref{derv:number-var} for the variance is verified with MC simulations in Fig~\ref{fig:cumulantnumber}a,b, respectively for $k=1.5$. We note that the variance scales with system size as $N^{\nu_k}$ with $\nu_k = (2-k)/(k+2)$. This implies that for $1<k<2$ the variance increases with increasing system size. For $k \to 2$ we see $\nu_k \to 0$ as a result $N^{\nu_k} \to \log(N)$, hence one generally expects that the variance grows logarithmically with $N$ similar to the case of the Dyson's log-gas ($k\to0$). For $k>2$, $\nu_k<0$ and the variance decreases with system size which suggests that the system becomes very rigid in the thermodynamic limit and possibly the typical fluctuations are dominated by microscopic fluctuations at the edges of the box. This is not captured by the present scaling analysis.\\

\noindent
\textit{Generalization to other quantities:} Using the same approach, we also study a more general quantity known as linear statistics defined as $S_N = \sum_{i=1}^N r(y_i)$, where $y_i=x_i/N^{\alpha_k}$ and $r(y)$ is an arbitrary function. The mean of this quantity scales linearly with system size as expected, whereas the variance scales as a power-law $\sim O(N^{\nu_k})$ with $\nu_k = (2-k)/(k+2)$ as described in Section.~\ref{sec5}. Note that the number distribution $\mathcal{N}(W,N)$ is also a linear statistic with the choice $r(y)=\Theta(w-y)\Theta(w+y)$, where $\Theta(x)$ is the Heavyside Theta function. Another interesting and well-studied quantity is the index defined as the number of particles, denoted by $\mathcal{I}(W, N)$, in the semi-infinite box $(-\infty, W]$ which corresponds to the choice $r(y) = \Theta(w-y)$ in the linear statistics. This quantity appears naturally in the study of the stability of complex systems~\cite{may1972will, wales2000energy}. It has been well studied in the context of the random matrix theory \cite{index1, castillo2016large}, the Dyson's log-gas~\cite{index2} and the 1dOCP model~\cite{rojas2018universal, dhar2018extreme}.  We find that the properties of the saddle point density profiles and the LDF corresponding to the index distributions are qualitatively similar to the number statistics problem summarized above. It is important to note that, in general, FCS behaves differently from the linear statistics with a smooth function $r(y)$~\cite{marino14, marino2016number, Flack2022, flack2022exact, Chen}. However, for the short-range case, this distinction does not seem to occur at least for the variance.

\section{Derivation of the number distribution}
\label{sec4}
In this section, we outline the derivation of the distribution of $\mathcal{N}(W, N)$, which quantifies the number of particles in the box $ [-W, W]$, as defined by
\begin{align}\label{numdef}
\mathcal{N}(W, N) = \sum_{i=1}^N \Theta(W-x_i)\Theta(W+x_i). 
\end{align}
Here $\Theta(x)$ is the Heaviside theta function. We start by writing the Gibbs-Boltzmann probability distribution of the position configuration in terms of the scaled variables $\{y_i =x_i/N^{\alpha_k}\}$ for $~i=1,2,\cdots,N$ [see Eq.~\eqref{scalexy}]: 
\begin{align}
    P\big(y_1,y_2,\cdots,y_N\big) = \frac{1}{Z_k(N, \beta)}~\exp\Big(-\beta N^{1+2\alpha_k} E_k\big(\{y_i\}\big)\Big),
\end{align}
where $E_k(\{y_i\})$ is the energy function in Eq.~\eqref{shamiltonian} and $Z_k(N, \beta)$ is the partition function, given by
\begin{align}\label{derv:partition-function}
    Z_k(N, \beta) = \int_{-\infty}^{\infty}~dy_1\int_{-\infty}^{\infty}~dy_2\ldots\int_{-\infty}^{\infty}~dy_N \exp\left(-\beta N^{1+2\alpha_k} E_k(\{y_i\})\right).
\end{align}
The mean of the number of particles can be easily computed as $ \langle \mathcal{N}(W, N) \rangle = \sum_i \langle \Theta(w-y_i)\Theta(y_i+w)\rangle,$ where $w = W/N^{\alpha_k}$ and $y_i=x_i/N^{\alpha_k}$. Simplifying further, we get
$\langle \mathcal{N}(W, N) \rangle \simeq c^*(w)~N$ where $c^*(w)$ is given in Eq.~\eqref{def:c^*}.

The distribution of $\mathcal{N}(W, N)$ can be obtained by integrating the microscopic configurations with the constraint of having $c\,N$ particles inside the box and it is given by 
\begin{align}\label{derv:prob-def}
	\mathscr{P}\left(\mathcal{N}= c~N, W\right) =  \int_{-\infty}^{\infty} dy_1 \ldots \int_{-\infty}^{\infty} dy_N~ &\frac{\exp{\left(-\beta N^{1+2\alpha_k} E_k\left(\{y_i\}\right)\right)}}{Z_{k}(N, \beta)}\times \notag \\& \delta\left(c~N - \sum_{i=1}^N \Theta(w+y_i)\Theta(w-y_i)\right),
\end{align}
where recall that $w = W/N^{\alpha_k}$. For the sake of brevity $W$ and $N$ in the argument of $\mathcal{N}(W, N)$ are suppressed.
For finite $N$, the integrals over the microscopic positions in Eqs.~\eqref{derv:partition-function} and~\eqref{derv:prob-def} are difficult to carry out for arbitrary $k$ with the only notable exceptions for $k \to 0$~\cite{Mehtabook} and $k=-1$~\cite{dhar2017exact, Flack2022, flack2022exact}.  However, in the large $N$ limit and for $\beta \sim O(1)$, the multiple integrals can be computed approximately using the Laplace method in which one first rewrites the microscopic integral as a path integral over density field configurations and then performs the saddle point calculation. Such a method in the literature is known as the Coulomb gas method \cite{dean06, dean08}, which has been used recently in the context of Riesz gases ~\cite{agarwal2019harmonically}. To adopt this field-theoretic method, we first rewrite the integral in Eq.~\eqref{derv:prob-def} as a path integral over the empirical density field 
$\rho(y) = \frac{1}{N}\sum_{i=1}^N \delta(y-y_i)$. More precisely, we compute the integral in Eq.~\eqref{derv:prob-def} in two steps: (i) we integrate over the microscopic positions corresponding to a density field $\rho(y)$ (which one could assume to be a smooth function in the large $N$ limit) and (ii) perform the integration over these density profiles. After the first step, one generates an entropy term $\mathcal{S}[\rho(y)]$ in the exponential in addition to the energy functional $\mathcal{E}_k[\rho(y)]$ to arrive at~\cite{agarwal2019harmonically}
\begin{align}\label{derv:prob-def3}
	\mathscr{P}\left(\mathcal{N}= c~N, W\right) =  &\int\mathcal{D}\left[ \rho(y) \right] \frac{\exp{\left(-\beta N^{1+2\alpha_k} \mathcal{E}_k\left[\rho(y)\right] + N\mathcal{S}\left[\rho(y)\right]\right)}}{Z_k(N, \beta)}\times \notag \\ &\delta\left(c~N-N\int_{-\infty}^{\infty} dy~\rho(y) \Theta(w+y)\Theta(w-y)\right) \delta\left(\int_{-\infty}^{\infty} dy~\rho(y)-1\right),
\end{align}
where the energy functional $\mathcal{E}_k\left[\rho(y)\right]$ is given in Eq.~\eqref{E-func-k>1}  and the entropy functional 
is given by~\cite{dean06, dean08}
\begin{align}\label{derv:jacobian}
 	\mathcal{S}[\rho(y)] = \left(-\int_{-\infty}^{\infty}dy~\rho(y)\log\rho(y)\right).
\end{align}
Using the integral representation of the delta function on the complex plane, one can  express $\mathscr{P}\left(\mathcal{N}, W\right)$ in Eq.~\eqref{derv:prob-def3} as
\begin{align}\label{derv:prob-def4}
	\mathscr{P}\left(\mathcal{N} = c~N, W\right) =  \int~d\mu \int~d\bar{\mu} \int\mathcal{D}\left[ \rho(y) \right] \frac{\exp{\left(-\beta~N^{1+2\alpha_k}G[\rho(y)]\right)}}{Z_k(N, \beta)},
\end{align}
with the action given by
\begin{align}\label{derv:action}
	G[\rho(y)]&= 	\mathcal{E}_k\left[\rho(y)\right] -\frac{T}{N^{2\alpha_k}} \mathcal{S}\left[\rho(y)\right]\notag \\&- \mu\left(\int_{-\infty}^{\infty}~dy~\rho(y)\big(1-\Theta(w+y)\Theta(w-y)\big)-1+c\right)\notag \\&- \bar{\mu}\left(\int_{-\infty}^{\infty}~dy~\rho(y)\Theta(w+y)\Theta(w-y)-c\right),
\end{align}
where $w = W/N^{\alpha_k}$. The functional $G[\rho(y)]$ in the above equation is essentially the free energy required to create a particular density profile with the chemical potentials $\bar{\mu}$ and $\mu$ ensuring that the fraction of particles inside and outside of the box $[-w,w]$ is  $c$ and $1-c$ respectively. 

Note that the factor $N^{1+2\alpha_k}$ in the exponent of Eq.~\eqref{derv:prob-def4} diverges for $N \to \infty$, since $1+2\alpha_k>0$ [see Eq.~\eqref{L_N-alpha_k}]. Therefore, the integral can be evaluated by a saddle point technique in which one needs to minimize the action in Eq.~\eqref{derv:action} w.r.t. the density field $\rho(y)$ as well as the chemical potentials $\bar{\mu}$ and $\mu$. Moreover, for large $N$ and $T \sim O(1)$ one can neglect the contribution from the entropy term in the saddle point calculation. We find the following equations
\begin{align}\label{derv:chem-chi}
	\bar{\mu}^* &= \frac{y^2}{2} + J \zeta(k)(k+1)\left(\varrho^{*}(y)\right)^{k}~~\text{for}~~|y|<w,\\
	\label{derv:chem-notchi}
	\mu^* &= \frac{y^2}{2} + J \zeta(k)(k+1)\left(\varrho^{*}(y)\right)^{k}~~\text{for}~~|y|>w,
\end{align}
along with the normalization constraints
\begin{align}
	\label{derv:norm-chi}
	\int_{-\infty}^{\infty}~dy~\varrho^*(y)\Theta(w+y)\Theta(w-y) &= c,\\
	\label{derv:norm-notchi}
	\int_{-\infty}^{\infty}~dy~\varrho^*(y)\Big(1-\Theta(w+y)\Theta(w-y)\Big) &= 1-c.
\end{align}
Here the ${}^*$ represents the saddle point values. Note that the saddle point equations in Eqs.~\eqref{derv:chem-chi} and~\eqref{derv:chem-notchi} are valid when the size ($2w$) of the box $[-w,w]$ is much larger than the (typical) mean inter-particle scaled distance {\it i.e.}, $w\gg O(1/N)$.

Note that the chemical potentials $\bar{\mu}^*$ and $\mu^*$ in Eqs.~\eqref{derv:chem-chi} and ~\eqref{derv:chem-notchi} are independent of the position $y$. On the other hand, the right-hand sides of Eqs.~\eqref{derv:chem-chi} and ~\eqref{derv:chem-notchi} diverge in the limit $y\to \infty$. This suggests that the saddle point density has a finite support. The density profile takes the form
\begin{align}
	\label{derv:rho}
	\varrho^*(y) = 
        \begin{cases}
        A_k \left(\bar{l}^2 - y^2\right)^{\frac{1}{k}}~~&\text{for}~ |y| \le {\rm  min}(w, \bar{l})
         \\ 
         A_k \left(l^2 - y^2\right)^{\frac{1}{k}}~~&\text{for}~ w \le |y| \leq l\\
        ~~~~~0~~~~&\text{otherwise}
        \end{cases},~
        {\text{where}}~\bar{l}=\sqrt{2 \bar{\mu}^*},~~l = \sqrt{2 \mu^*},
\end{align}
and the constant $A_k$ is given in Eq.~\eqref{A_1k}. The length scales $\bar{l} \equiv {\bar l}(c, w)$ and $l \equiv l(c,w)$ are functions of the two parameters $c$ and $w$ and are obtained from the normalization conditions  Eqs.~\eqref{derv:norm-chi} and~\eqref{derv:norm-notchi}, respectively. We find 
\begin{align}
	\label{derv:number-suppq}
	\bar{l} &=  \begin{cases}
    c^{\alpha_k} l_0~&\text{for}~c\leq\bar{c}(w),\\
    c^{\alpha_k} l_0 I\left(\left(\dfrac{w}{\bar{l}}\right)^2, \frac{1}{2}, \frac{1}{k}+1\right)^{-\alpha_k}~&\text{for}~c\geq\bar{c}(w)
    \end{cases},
\end{align}
 where $\bar{c}(w)$ is the fraction for which $\bar{l}=w$ which is given by
\begin{align}\label{cbar}
    \bar{c}(w) = \left(\frac{w}{l_0}\right)^{\frac{1}{\alpha_k}}.
\end{align}
In Eq.~\eqref{derv:number-suppq} we have introduced the function $I(h,a,b)$ which is defined as
\begin{align}\label{I(g,a,b)-def}
I(h,a,b) = \frac{B(h,a,b)}{B(1, a,b)}~\text{with}~B(h,a,b) = \int_{0}^{h}ds~s^{a-1}(1-s)^{b-1}.
\end{align}
For the edge $l$ of the density profile outside the box, we find 
 \begin{align}\label{derv:number-suppm}
	l &= (1-c)^{\alpha_k} l_0 \Bigg(1-I\bigg(\left(\frac{w}{l}\right)^2, \frac{1}{2}, \frac{1}{k}+1\bigg)\Bigg)^{-\alpha_k}.
\end{align}
We can now numerically compute the lengths $l$ and $\bar{l}$ for any $c$ and $w$ by solving the transcendental Eqs.~\eqref{derv:number-suppq} and~\eqref{derv:number-suppm}. Note that the edge of the support of the density profile inside the box is ${\rm min}(w, \bar{l})$. The extent of this support depends on the fraction of particles $c$ in the box $[-w, w]$. For $c$ less than a certain value $\bar{c}(w)$ we find $\bar{l}<w$ while for $c>\bar{c}(w)$ we get $\bar{l}>w$. As $c$ is changed the shape of the density profile changes and we obtain three distinct regimes (as depicted in Fig.~\ref{fig:phase}) namely: (I) low (II) moderate and (III) high fraction regimes. We further elaborate on these regimes below.  

\begin{figure}[htb]
\centering
\includegraphics[scale=0.33]{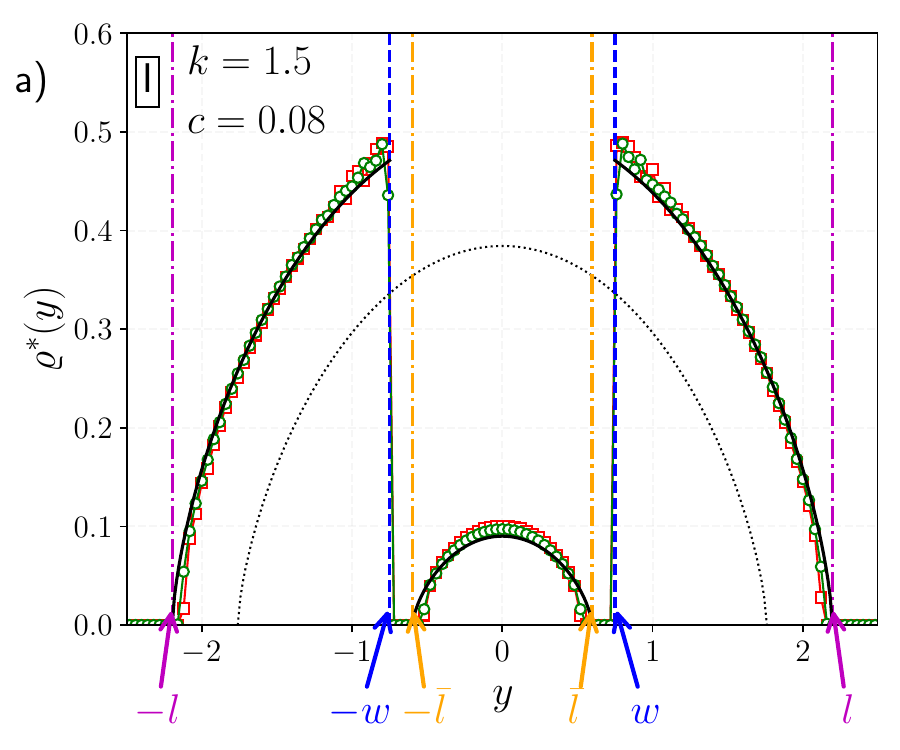}
\includegraphics[scale=0.33]{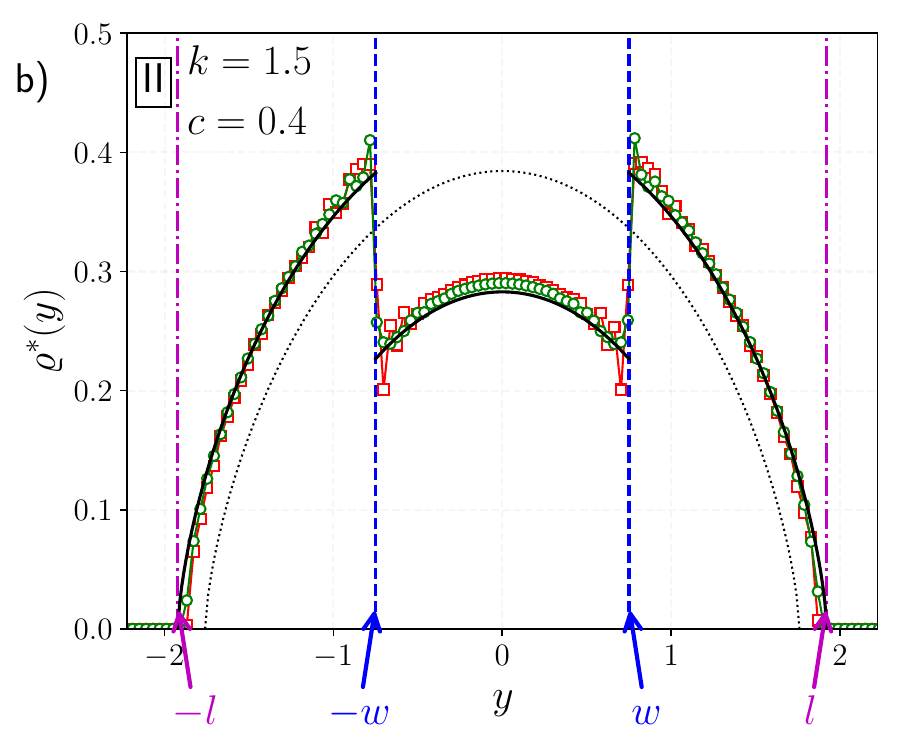}
\includegraphics[scale=0.33]{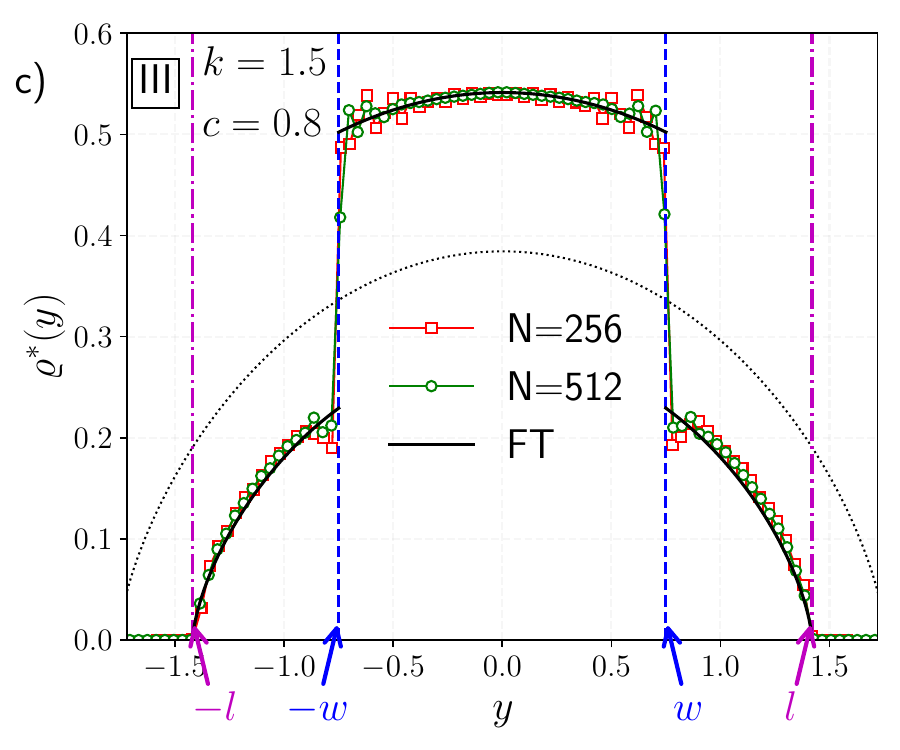}
\caption{Plots of the equilibrium density profiles for different fractions of particles $c$ confined inside the box $[-w,w]$ with $w=0.75$, $k=1.5$ and $J=1$.  The boundaries of the box at $y=\pm w$ are shown by the blue vertical dashed lines.  We consider three values of $c$, (a) $0.08$ (b) $0.4$ and (c) $0.8$ which are representative of the three regimes: (I) $c<\bar{c}(w)$, (II) $\bar{c}(w)<c<c^*(w)$ and (III) $c^*(w)<c$, respectively, where $c^*(w)$ is given in Eq.~\eqref{def:c^*} and $\bar{c}(w)$ is given in Eq.~\eqref{cbar}. For $w = 0.75$ and $k=1.5$ one finds that $\bar{c}(w) = 0.136$ and $c^*(w) = 0.552$. The symbols in all the plots are obtained using MC simulation for $N=256$ and $N=512$ whereas the solid lines represent the theoretical results given Eq.~\eqref{derv:rho}. The dotted line in each plot represents the density profiles $\rho_0(y)$ [see Eq.~\eqref{rho_uc}] in the unconstrained case {\it i.e.}, without any wall, from which one can compute the fraction $c^*(w)$ of particles within the region $[-w,w]$. Here we have taken an average of over $10^6$ samples for all the plots.}
    \label{fig:densnumber}
\end{figure}

\textit{(I) Low fraction $[0\leq c<\bar{c}(w)]$:} As shown in Fig.~\ref{fig:densnumber}a, in this regime, due to the low fraction of the particles $c$ within the box, the density profile inside forms a small droplet at the minimum of the harmonic trap. It does not spread over the full extent of the box $[-w, w]$ and is only supported over the region $[-\bar{l}, \bar{l}]$. This leads to the appearance of two holes with no particles between the droplet and the edges of the box. Outside of the box, the remaining particles form truncated domes on both sides. The support of the left dome is  $[-l, -w]$ while for the right dome, it is $[w, l]$. As we further increase the fraction $c$, the edge of the support of the droplet $\bar{l}$ increases and eventually touches the edges of the box located at $\pm w$ when $c= \bar{c}(w)$. For the sake of brevity, we sometimes suppress the $w$ dependence in $\bar{c}(w)$.

\textit{(II) Moderate fraction $[\bar{c}(w)<c \leq c^*(w)]$:} As shown in Fig.~\ref{fig:densnumber}b, as the fraction $c$ is increased above $\bar{c}(w)$, the droplet grows but its support does not expand. As a consequence, the density at the walls just inside the box increases and this droplet becomes a truncated dome. Outside the box, the support of the left and right truncated domes shrink. The value of the density just outside the box decreases. Therefore the density profile is discontinuous at the locations of the wall ($\pm w$) [see Fig.~\ref{fig:densnumber}b and Fig.~\ref{fig:phase} (inset)]. As the fraction $c$ inside the box is further increased, the jump in the value of the density at the location of the wall is reduced. This jump eventually disappears when the fraction inside the box becomes the same as the fraction $c^*(w)$ [see Eq.~\eqref{def:c^*}]. Hence in this regime with $\bar{c}(w)<c<c^*(w)$, the density profile has three parts: two truncated domes on either side of the box and another truncated dome inside the box. 

\textit{(III) High fraction $[c^*(w)<c<1]$:} When $c>c^*(w)$, we find that the density at the wall just inside the box, increases further and becomes higher than that of the density at the wall just outside the box [see Fig.~\ref{fig:densnumber}c and Fig.~\ref{fig:phase} (inset)]. Therefore the density profile in this regime, with $c>c^*(w)$, comprises of three truncated domes. 

In Fig.~\ref{fig:densnumber}a,b,c, we plot the density profiles given in Eq.~\eqref{derv:rho} for the three regimes along with the same obtained from MC simulation and we observe an excellent agreement. The above discussion was based on varying $c$ with the wall position $w$ fixed. Similarly, one could obtain these three regimes by varying the wall position $w$ by keeping the fraction $c$ fixed [see Fig.~\ref{fig:phase}].

\begin{figure}[htb]
    \centering
    \includegraphics[scale=0.6]{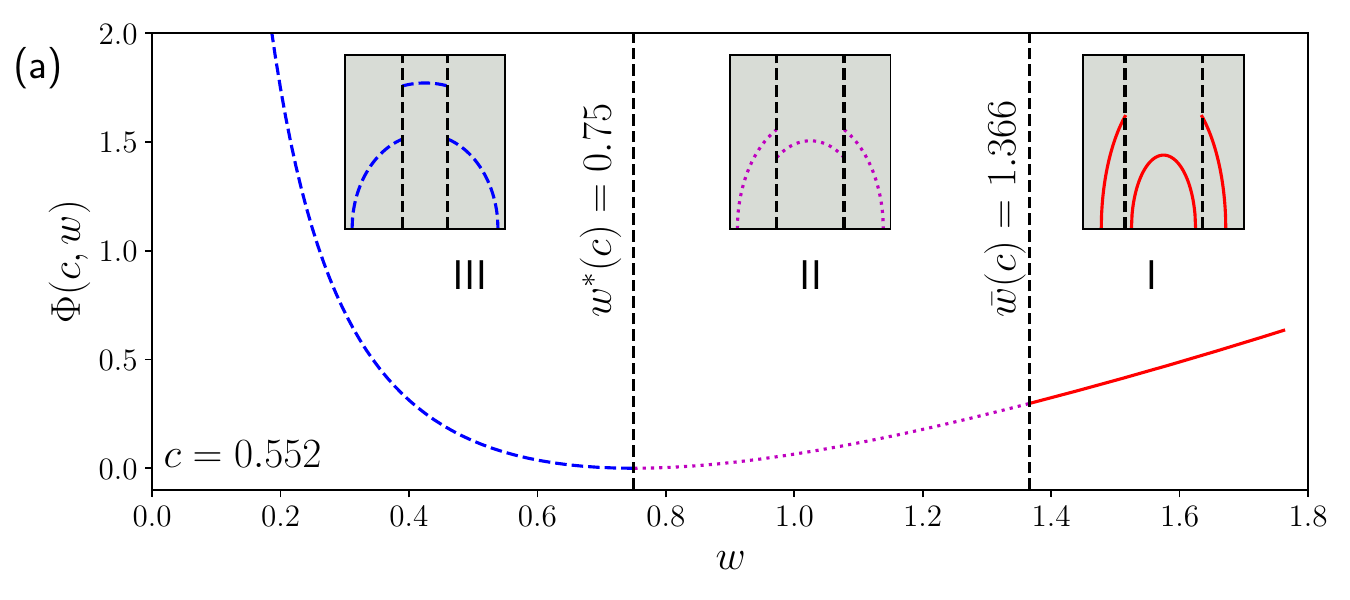}\\ 
    \includegraphics[scale=0.6]{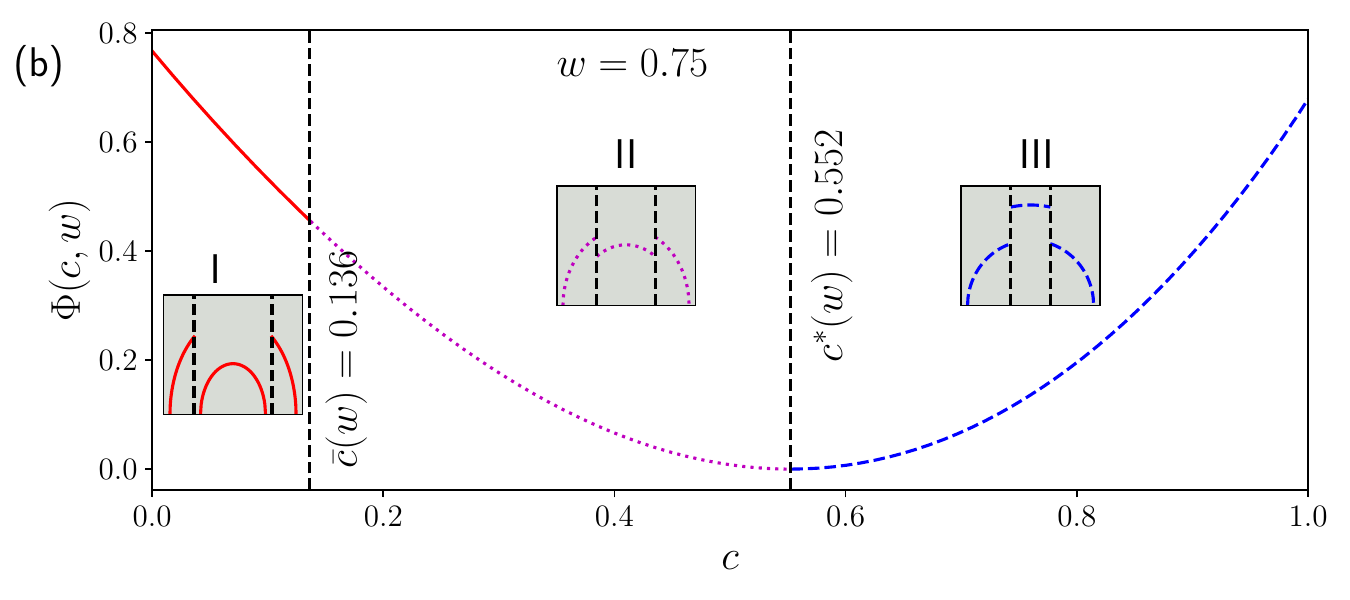}
    \caption{The plot displays the large deviation function $\Phi(c,w)$, given in Eq.~\eqref{derv:number-ldf1}, for $k=1.5$ and $J=1$. In (a) we show a plot of $\Phi(c,w)$ as a function of $w$ for $c=0.552$ and in (b) we plot $\Phi(c,w)$ as a function of $c$ for $w=0.75$. In both plots, we have demarcated three regions (I, II and III) based on the three types of saddle point density profiles (as shown in Fig.~\ref{fig:densnumber}) that create the large deviations in those three different regions.  Schematic plots of such density profiles are provided in the insets.  }
    \label{fig:phinumber}
\end{figure}

As a next step in computing the integral in Eq.~\eqref{derv:prob-def4}, we substitute the saddle point density profile from Eq.~\eqref{derv:rho} in the expression of the action, $G[\varrho^*(y)]$, in Eq.~\eqref{derv:action}, we find the following large deviation form for the probability distribution
\begin{align}\label{derv:prob-def5}
	\mathscr{P}\left(\mathcal{N} = c~N, W\right) \asymp  \exp\Big(-\beta N^{1+2\alpha_k} \Phi(c,w)\Big),~\text{with}~w = \frac{W}{N^{\alpha_k}}
\end{align}
where the large deviation function is given by
\begin{align}\label{derv:ldf}
   \Phi(c,w) = G\left[\varrho^*(y)\right] - G_0.
\end{align}
Here the unconstrained action, $G_0$, is given by the logarithm of the partition function in Eq.~\eqref{derv:partition-function}. For large $N$ it reads ~\cite{jit2022} 
\begin{align}\label{derv:unconstrained-action}
	G_0 \approx -\frac{\log{Z_k(N, \beta)}}{\beta N^{1+2\alpha_k}} =  \frac{ l_0^{2}(k+2)}{2(3k+2)},
\end{align}
where $l_0$ is the edge of the support of the unconstrained density profile $\rho_0(y)$ [see  Eq.~\eqref{A_1k}]. By neglecting the contribution from the entropy term in Eq.~\eqref{derv:action}, we approximate the action by
 \begin{align}\label{derv:action1}
	G&\left[\varrho^*(y)\right] = \mathcal{E}_k[\varrho^*(y)] 
\end{align}
where the energy functional is given in Eq.~\eqref{E-func-k>1}. After simplifying Eq.~\eqref{derv:ldf} we obtain the LDF as 
\begin{align}\label{derv:number-ldf1}
	\Phi(c, w) =   \begin{cases}
    G_0\Bigg((1-c)^{\frac{3k+2}{k+2}}\mathscr{H}\left(\frac{w}{l}\right)+(c)^{\frac{3k+2}{k+2}}-1\Bigg),~&\text{for}~c\leq\bar{c}(w),\\
    G_0\Bigg((1-c)^{\frac{3k+2}{k+2}}\mathscr{H}\left(\frac{w}{l}\right)+(c)^{\frac{3k+2}{k+2}}\mathscr{J}\left(\dfrac{w}{\bar{l}}\right)-1\Bigg),~&\text{for}~c\geq\bar{c}(w).
  \end{cases}
\end{align}
The length scales $\bar{l}$ and $l$ are given in Eqs.~\eqref{derv:number-suppq} and~\eqref{derv:number-suppm}. The functions $ \mathscr{H}(h)$ and $\mathscr{J}(h)$ in Eq.~\eqref{derv:number-ldf1} are simple and given by 
\begin{align}
    \mathscr{H}(h) &= \left(1-I\left(h^2,\frac{1}{2},1+\frac{1}{k}\right)\right)^{-\frac{2k}{k+2}} \notag\\&+ \frac{h (1-h^2)^{\frac{1}{k}+1} \left(2  k^2\right)}{(k+1) (k+2) B\left(\frac{1}{2},1+\frac{1}{k}\right)}\left(1-I\left(h^2,\frac{1}{2},1+\frac{1}{k}\right)\right)^{-\frac{3k+2}{k+2}},\label{derv:number-h}\\
    \mathscr{J}(h) &=\left(I\left(h^2,\frac{1}{2},1+\frac{1}{k}\right)\right)^{-\frac{2k}{k+2}} \notag\\&- \frac{h (1-h^2)^{\frac{1}{k}+1} \left(2  k^2\right)}{(k+1) (k+2) B\left(\frac{1}{2},1+\frac{1}{k}\right)}\left(I\left(h^2,\frac{1}{2},1+\frac{1}{k}\right)\right)^{-\frac{3k+2}{k+2}},\label{derv:number-j}
\end{align}
where $I(h,a,b)$ is given in Eq.~\eqref{I(g,a,b)-def}.  In Fig.~\ref{fig:phinumber}a. we show the LDF $\Phi(c,w)$ given in Eq.~\eqref{derv:number-ldf1} as a function of $w$ (for fixed $c$) and in Fig.~\ref{fig:phinumber}b we show the variation of LDF with $c$ (for fixed $w$). The three types of saddle point density profiles corresponding to the three regions I, II, and III, shown in Fig.~\ref{fig:densnumber}, determine the form of the LDF as shown in Fig.~\ref{fig:phinumber}a,b. Next, we discuss the asymptotic behaviour of $\Phi(c,w)$ in different limits.

\begin{figure}[htb]
    \centering
    \includegraphics[scale=0.6]{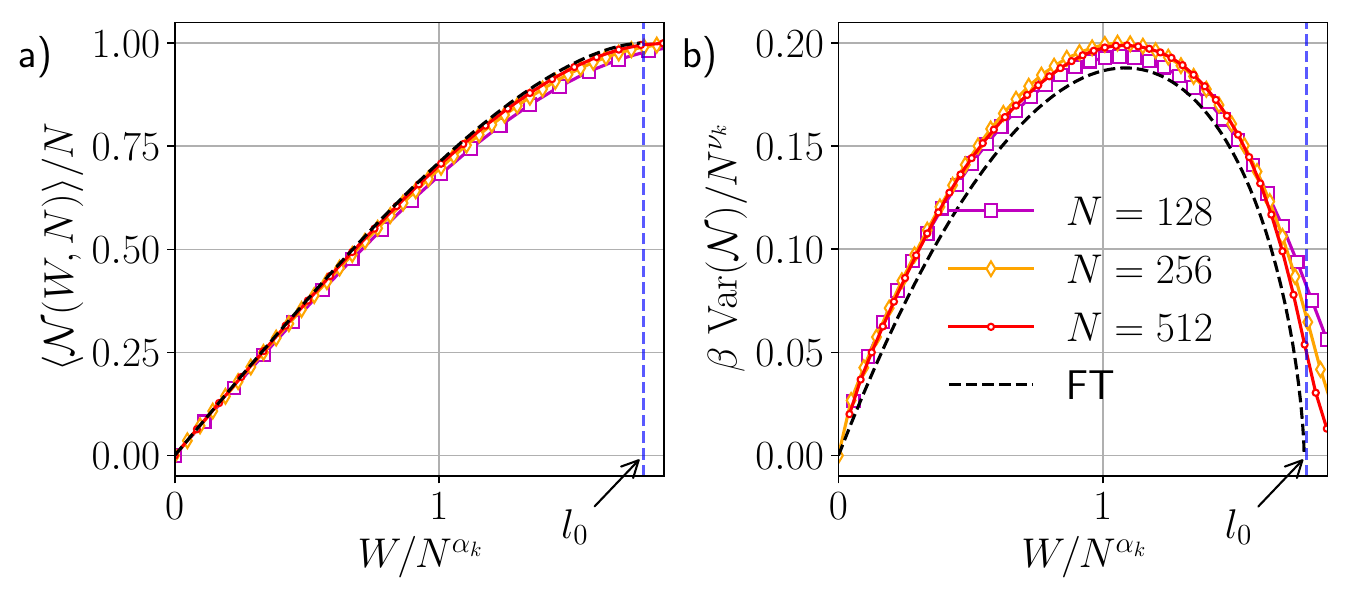}
    \caption{Numerical verification of the mean and variance of the number distribution problem. The plot displays the $W$ dependence of (a) the mean fraction of particles and (b) the scaled variance of the number of particles in the box $[-W,W]$ for $N=128, 256, 512$ with parameters $k=1.5$, $T=10$ and $J=1$. The vertical blue dashed line represents the box with $W = l_0 N^{\alpha_k}$. The symbols indicate the results obtained from the Monte Carlo simulations and they are compared with our theoretical predictions (solid lines) given by Eqs.~\eqref{cstarn0} and ~\eqref{derv:number-var-m} for the mean and the variance, respectively. 
    }
    \label{fig:cumulantnumber}
\end{figure}

\textit{Behaviour around $c^*(w)$:}  We start with the behaviour of $\Phi(c,w)$ near $c^*(w)$ which describes the probability of typical fluctuations of $c$ around $c^*(w)$. Recall from Eq.~\eqref{def:c^*} that $c$ represents the mean fraction of particles inside a box of size $w$. Setting $c = c^*(w) + \kappa$ in Eq.~\eqref{derv:number-ldf1} and expanding to leading order in $\kappa$ we find $\Phi(c^*(w)+\kappa,w) \propto \kappa^2$. 
A slightly different derivation of this expansion is given in 
Appendix.~\ref{appendlinear}. The quadratic behaviour of $\Phi(c^*(w)+\kappa,w)$ with $\kappa$ in the leading order implies a Gaussian distribution for the typical fluctuations  given by 
\begin{align}\label{derv:number-prob-typical}
	\mathscr{P}\left(\mathcal{N} = (c^*(w)  + \kappa)N,W\right) \asymp  \exp\Bigg(-\frac{N^2\kappa^2}{2~{\rm Var}(\mathcal{N})}\Bigg),
\end{align}
with the variance of the number of particles given by
\begin{align}
	{\rm Var}(\mathcal{N}) &= \frac{N^{\nu_k}}{\beta~l_0^{2}~\alpha_k} \mathcal{V}\left(\frac{W}{N^{\alpha_k}l_0} \right),~
 \text{with}~\nu_k = 1-2\alpha_k = \frac{2-k}{k+2},
 \label{derv:number-var-m} \\
 \text{and}~&~
 \mathcal{V}(h) = I\left(h^2,\frac{1}{2},\frac{1}{k}\right) \left(1- I\left(h^2,\frac{1}{2},\frac{1}{k}\right)\right),\label{mcal(N)_1} 
\end{align}
where the function $I(h,a,b)$ is given in Eq.~\eqref{I(g,a,b)-def}. For small $h$, the function $\mathcal{V}(h) \propto h$ whereas for $h \to 1$, $\mathcal{V}(h) \propto (1-h)^{1/k}$.  In Fig.~\ref{fig:cumulantnumber}a,b, we compare our theoretical results for the mean [see Eq.~\eqref{def:c^*}] and variance [see Eq.~\eqref{derv:number-var-m}] of $\mathcal{N}$  with the same measured in MC simulations and observe good agreement everywhere except the edges where it matches better for large-$N$. This is due to finite-$N$ corrections. Note that for $k>2$ the exponent $\nu_k$, given in Eq.~\eqref{derv:number-var-m}, is negative implying that the variance decreases with increasing system size. This suggests that the contribution to the typical fluctuations for large-$N$ are primarily due to the microscopic fluctuation at the edges of the box. Such fluctuations do not cause changes in the density profile and are thereby missed in the field theory description. In the marginal case of $k=2$, as mentioned earlier, the exponent $\nu_k=0$ possibly suggests  $\log(N)$ growth of the variance.

\begin{figure}
    \centering
    \includegraphics[scale=0.7]{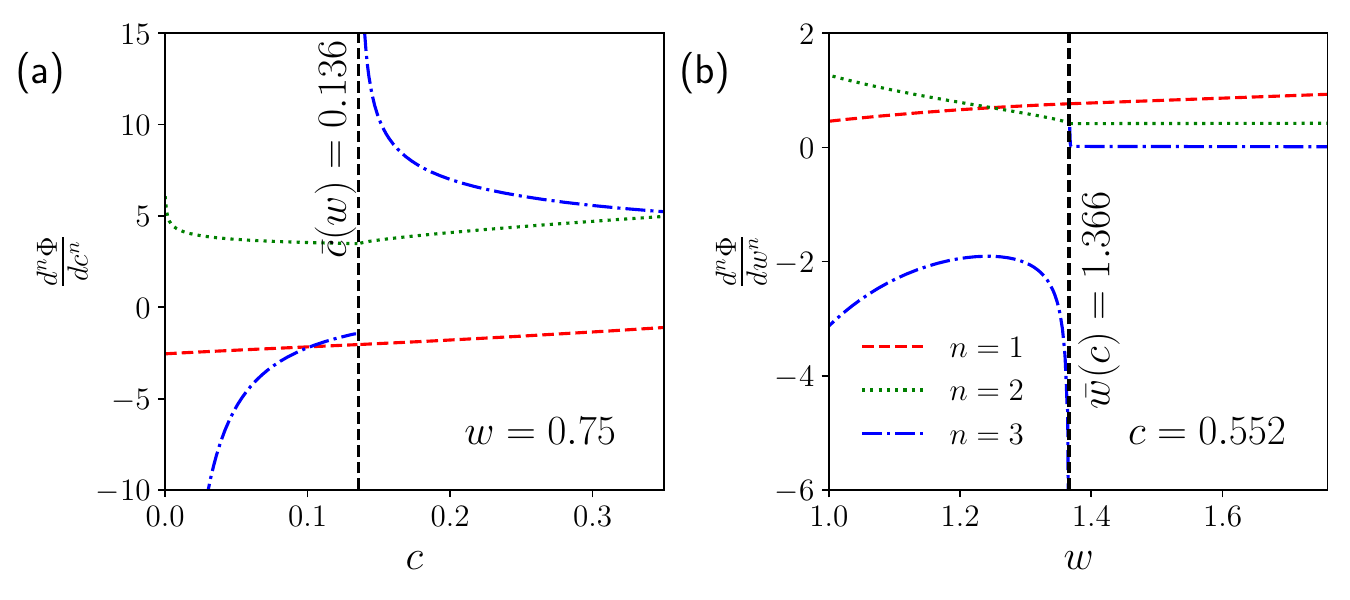}
    \caption{The plot displays the derivatives of the LDF $\Phi(c,w)$ [Eq.~\eqref{derv:number-ldf1}]. Specifically, it shows the first (red dashed line), second (green dotted line) and third (blue dash-dotted line) derivatives: (a) w.r.t. $c$ for $w=0.75$ and (b) w.r.t. $w$ for $c=0.552$. Notably, we observe a pronounced discontinuity in the third derivative of $\Phi(c,w)$ in (a) at $c=\bar{c}(w)$ [see Eq.~\eqref{cbar}] and in (b) at $w=\bar{w}(c) = l_0 c^{\alpha_k}$, establishing the presence of third order phase transitions.}
    \label{fig:3dphin}
\end{figure}

\textit{Non-analytic behaviour and phase transitions:} We recall that $\bar{c}(w)$ given in Eq.~\eqref{cbar} represents the fraction at which the hole in the density profile inside the box vanishes. We find that this hole-closing phenomenon gives rise to non-analytic properties of $\Phi(c,w)$ around $c=\bar{c}(w)$. Expanding the LDF $\Phi\big(\bar{c}(w)(1+\epsilon),w\big)$ for small $\epsilon$ we find [see Appendix~\ref{phivsc}]:
\begin{align}
    \Phi\big(\bar{c}(w)(1+\epsilon),w\big)-\Phi\big(\bar{c}(w),w\big) &= 
    \begin{cases}
       \tilde{\mathbb{A}}~\epsilon + \tilde{\mathbb{B}}~\epsilon^2 + \tilde{\mathbb{C}}_{+}~\epsilon^{2+\frac{1}{k}} + o(\epsilon^{2+\frac{1}{k}}),&{\rm for}~\epsilon >0\\
        \tilde{\mathbb{A}}~\epsilon + \tilde{\mathbb{B}}~\epsilon^2 + \tilde{\mathbb{C}}_{-}~\epsilon^3+ O(\epsilon^{4}),&{\rm for}~\epsilon <0
    \end{cases},
   \label{phisimc}
\end{align}
where the constants $\tilde{\mathbb{A}}$, $\tilde{\mathbb{B}}$ etc. are given in the Eqs.~\eqref{atilde}-\eqref{ctilde-} of the Appendix~\ref{phivsc}. Note that $o(\epsilon^{2+\frac{1}{k}})=O(\min[\epsilon^{2+\frac{2}{k}},\epsilon^3])$. For a fixed $w$, the third derivative of the LDF $\Phi(c,w)$ w.r.t. $c$ ({\it i.e.}, $\epsilon$) shows a discontinuity at $c=\bar{c}(w)$ [Eq.~\eqref{cbar}] as demonstrated in Fig.~\ref{fig:3dphin}a. More precisely, the third derivative is finite for $c \to \bar{c}^{\,-}(w)$ and is diverging for $c \to \bar{c}^{\,+}(w)$. Similar discontinuities in the third derivative of the LDF have been observed previously in various other contexts and have been associated to third order phase transition -- such as linear statistics in $1$d Coulomb gas \cite{Flack2021} and in extreme statistics of Riesz gas \cite{jit2022}, of Coulomb gas  \cite{cunden2018universality} and random matrix theory \cite{majumdar2014top, Cunden_s}. The non-analyticity of the  LDF, described above, stems from the structural change of the saddle point density profile from (I) Low to (II) Moderate fraction regime.

By the same mechanism, a similar third-order phase transition is expected to occur when we cross the line $c = \bar{c}(w)$ horizontally in Fig.~\eqref{fig:phase} {\it i.e.}, by varying $w$ while keeping $c$ fixed as demonstrated in Fig.~\ref{fig:3dphin}b. This phase transition can be shown by analyzing the behaviour of $\Phi(c,w)$ near the special box size $\bar{W}=\bar{w}(c)N^{\alpha_k}$ with $\bar{w}(c) = l_0\,c^{\alpha_k}$. While reducing the box size starting from a larger value, the hole in the density profile inside the box (containing $c~N$ particles) decreases and at a special value $\bar{w}(c)$ of the box size, the droplet touches the boundaries of the box. In Appendix~\ref{phivsw} we show that the LDF $\Phi(c,w)$ around $\bar{w}(c)$ also exhibits non-analytic behaviour. More elaborately it behaves as
\begin{align}\label{phisimw}
    \Phi(c,\bar{w}(c)(1+\epsilon))-\Phi(c,\bar{w}(c)) &= 
    \begin{cases}
    \tilde{\mathbb{D}}~\epsilon + \tilde{\mathbb{E}}~\epsilon^2 + \tilde{\mathbb{F}}_{+}~\epsilon^3 +O(\epsilon^{4}), & \text{for}~\epsilon>0\\
    \tilde{\mathbb{D}}~\epsilon + \tilde{\mathbb{E}}~\epsilon^2 + \tilde{\mathbb{F}}_{-}~|\epsilon|^{2+\frac{1}{k}} + o(\epsilon^{2+\frac{1}{k}}), &\text{for}~\epsilon<0       
    \end{cases},
\end{align}
where the constants $\tilde{\mathbb{D}}$, $\tilde{\mathbb{E}}$ etc. are given in the Eqs.~\eqref{etilde}-\eqref{ftilde-} of the Appendix~\ref{phivsw}. We find that the third derivative is discontinuous at $w=\bar{w}(c)$. More precisely, $\Phi(c,w)$ is finite for $w \to \bar{w}(c)_+$ and is diverging for $w \to \bar{w}(c)_-$.

\noindent 
\textit{Behaviour of $\Phi(c, w)$ near $c=0$ and $c=1$:}  In these limits we find the following approximations for a given $w$,
\begin{align}
\Phi(c, w) \approx &~\Phi(0, w) - \mu^* c,~~~~~~~~~~\text{for}~~c\to0 \label{derv:ldfnepsi},\\
\Phi(c, w) \approx &~\Phi(1, w) - \bar{\mu}^*(1-c)~~~~~\text{for}~~c\to1 \label{derv:ldfn1epsi}.
\end{align}
Here, $\bar{\mu}^*$ and $\mu^*$ represent the chemical potentials [see Eqs.~\eqref{derv:chem-chi} and~\eqref{derv:chem-notchi}] of the gas outside and inside the box for $c=0$ and $c=1$, respectively. Here the LDFs $\Phi(0,w)$ and $\Phi(1,w)$ describe the probability of having no particle and all the particles in the box.  The linear dependence on $c$ in Eq.~\eqref{derv:ldfnepsi} represents the energy cost for depositing fraction $c$ of particles into the initially empty box. On the other hand, $ -\bar{\mu}^*(1-c)$ represents the energy cost for taking out $(1-c)$ of particles out of an initially filled box.

\noindent \textit{Hole formation LDF $\Phi(0, w)$:} Taking $c=0$ in Eq.~\eqref{derv:number-ldf1}, we find that the hole formation LDF $\Phi(0,w)$ is given by 
\begin{align}\label{derv:number-ldf2}
	\Phi(0, w) = G_0 \times \left(\mathscr{H}\left(\frac{w}{l}\right)-1\right),
\end{align}
where $G_0 $ is given in Eq.~\eqref{derv:unconstrained-action}, $\mathscr{H}(h)$ is given by Eq.~\eqref{derv:number-h}, and $l$ is the edge of the support of the (scaled) density profile when the box $[-W,W]$ is empty. Numerically, $l$ can be calculated using Eq.~\eqref{derv:number-suppm} for $c=0$. 

\noindent \textit{Complete confinement LDF $\Phi(1, w)$:} Taking $c=1$ in Eq.~\eqref{derv:number-ldf1}, we find the LDF associated to the probability of containing all the particles in the box. This LDF is given by
\begin{align}\label{derv:number-ldf3}
	\Phi(1, w) = G_0 \times \left(\mathscr{J}\left(\dfrac{w}{\bar{l}}\right)-1\right),
\end{align}
where $G_0 $ is given in Eq.~\eqref{derv:unconstrained-action}, $\mathscr{J}(h)$ is given by Eq.~\eqref{derv:number-j}, and $\bar{l}$ represents the edge of the support of the density profile when the box contains $N$ particles. Numerically, $\bar{l}$ can be calculated using Eq.~\eqref{derv:number-suppq} for $c=1$.

\vspace{0.5cm}
\noindent
\section{Index distribution}\label{indist}
Another interesting observable is the index for which our calculation presented in Section~\ref{sec4} for studying the number distribution can be straightforwardly extended. The index denoted by $\mathcal{I}(W, N)$, counts the number of particles below a certain position $W$ and it is defined as 
\begin{align}\label{indef}
    \mathcal{I}(W, N) = \sum_i^{N}\Theta(W-x_i),
\end{align}
where $x_i$ is the position of the $i^{\rm th}$ particle for any $i \in {1,2,\ldots,N}$ and $\Theta(x)=1$ for $x\geq0$ and zero otherwise. Here we focus on $W>0$. The case with $W<0$  can be obtained from the distribution with $W>0$ using the relation
\begin{align}
\text{Prob.}\left[ \mathcal{I}(W, N)=I_d\right] & = \text{Prob.}\left[ \mathbb{N}_{[W,\infty)}(N)=N-I_d\right]
= \text{Prob.}\left[ \mathcal{I}(-W, N)=N-I_d\right],
\end{align}
where $\mathbb{N}_{[W, \infty)}(N) = \sum_{i=1}^N \Theta(x_i-W)$ represents the number of particles to the right of $W$. To obtain the second equality we have used the inversion symmetry of the energy function $\tilde{E}_k(\{x_i\}) = \tilde{E}_k(\{-x_i\})$. To find the probability distribution of $\mathcal{I}(W,N)$, we follow the same procedure as done in the previous section. We find that for large $N$ this probability distribution has the following large deviation form
\begin{align}\label{derv:index-prob}
	\mathcal{P}\left(\mathcal{I} = c~N,W\right) \asymp  \exp\Bigg(-\beta N^{1+2\alpha_k}\Psi(c, W/N^{\alpha_k})\Bigg),
\end{align}
where $\Psi(c,w)$ is the LDF and $\alpha_k = k/(k+2)$. We find that the LDF is given by
\begin{align}\label{ldfindex}
    \Psi(c,w) = \begin{cases}
    \frac{G_0}{2}\left((2c)^{2\alpha_k+1} + (2(1-c))^{2\alpha_k+1}\mathscr{H}\Big(\frac{w}{l}\Big)-2\right),&~\text{for}~c\leq\bar{c}(w)\\
    \frac{G_0}{2}\left((2c)^{2\alpha_k+1}\mathscr{J}\Big(\dfrac{w}{\bar{l}}\Big) + (2(1-c))^{2\alpha_k+1}\mathscr{H}\Big(\frac{w}{l}\Big)-2\right),&~\text{for}~c>\bar{c}(w)
    \end{cases},
\end{align}
where $G_0 $ is given in Eq.~\eqref{derv:unconstrained-action}, $w = W/ N^{\alpha_k}$, $\bar{c}(w)=\left(\tfrac{w}{l_0}\right)^{\frac{1}{\alpha_k}}$ and the functions $\mathscr{J}(h)$ and $\mathscr{H}(h)$ are 
\begin{align}\label{appendix:ji}
    \mathscr{J}(h) &=  \Bigg(1+I\bigg(h^{2}, \frac{1}{2}, \frac{1}{k}+1\bigg)\Bigg)^{-2\alpha_k}\notag\\&-\frac{h\left(1-h^{2}\right)^{\frac{1}{k}+1}}{B\left(\frac{1}{2},1+\frac{1}{k}\right)}\frac{2k^2}{(k+1)(k+2)}\Bigg(1+I\bigg(h^{2}, \frac{1}{2}, \frac{1}{k}+1\bigg)\Bigg)^{-2\alpha_k-1},\\
    \label{appendix:hi}
    \mathscr{H}(h) = \Bigg(&1-I\left(h^{2}, \frac{1}{2}, \frac{1}{k}+1\right)\Bigg)^{-2\alpha_k} \notag\\&-\frac{h \left(1-h^{2}\right)^{\frac{1}{k}+1}}{B\left(\frac{1}{2},1+\frac{1}{k}\right)}\frac{2 k^2}{(k+1)(k+2)}\Bigg(1-I\bigg(h^{2}, \frac{1}{2}, \frac{1}{k}+1\bigg)\Bigg)^{-2\alpha_k-1}.
\end{align}
The length scales $\bar{l} \equiv \bar{l}(c,w)$ and $l \equiv l(c,w)$ in Eq.~\eqref{ldfindex} are functions of $c$ and $w$. These can be obtained by numerically solving the following transcendental equations
\begin{align}
	\label{derv:index-suppq}
	\bar{l}&=  \begin{cases}
    c^{\alpha_k} l_0~&\text{for}~c\leq\bar{c}(w),\\
    (2c)^{\alpha_k} l_0 \left(1+I\left(\left(\dfrac{w}{\bar{l}}\right)^2, \frac{1}{2}, \frac{1}{k}+1\right)\right)^{-\alpha_k}~&\text{for}~c\geq\bar{c}(w)
    \end{cases},
 \end{align}
\begin{align}\label{derv:index-suppm}
	l &= (2(1-c))^{\alpha_k} l_0 \Bigg(1-I\bigg(\left(\dfrac{w}{l}\right)^2, \frac{1}{2}, \frac{1}{k}+1\bigg)\Bigg)^{-\alpha_k},
\end{align}
where the function $I(h,a,b)$ is defined in Eq.~\eqref{I(g,a,b)-def}. We analyze the behaviour of the LDF $\Psi(c,w)$ at $c = \bar{c}(w)$ for a fixed $w$. We find that it shows a third-order phase transition, similar to the `number' problem. In this case also, the distribution of the typical fluctuations of $\mathcal{I}(W,N)$ is Gaussian distribution and the variance scales as $N^{\nu_k}$ with $\nu_k = (2-k)/(k+2)$.\\

\noindent \textit{Pressure and Bulk modulus:} Using the LDF for Index distribution, we can compute the thermodynamic pressure and the bulk modulus [see Appendix~\ref{pressureandbulkmodulus}]. 

\section{Linear statistics}
\label{sec5}
In this section, we study the linear statistics which is defined as
\begin{align}
S_N= \sum_{i=1}^N r(y_i),
\label{r_def}
\end{align}
where the function $r(y)$ is arbitrary and recall, from Eq.~\eqref{scalexy}, that $y_i = x_i/N^{\alpha_k}$ denote the scaled variables. Linear statistics generalizes FCS, for example, by choosing the function $r(y)$ appropriately we can obtain both the number and index distribution problems:
\begin{align}
r(y) =
\begin{cases}
 \Theta(y+w)\Theta(w-y)~&~{\rm Number~statistics} \\
\Theta(w-y) ~&~{\rm Index ~distribution}
\end{cases}
\end{align}
where $\Theta(y)$ is the Heaviside theta function. Linear statistics can be used to study the ground state properties of the system in arbitrary traps. It has been widely studied in both mathematics and physics~\cite{Politzer, Beenaker1993, Beenaker1993B, Basor, Chen, baker, Joh, Soshnikov, Pastur2, kanzieper1, bohigas1, sommers, Lytova, Khor, kanzieper2, bohigas2, kedar, Texier13, Cunden2014, Grabsch1, Cunden_s, Grabsch4, Grabsch2, Grabsch3, Grabsch5, valov2024large}. Interestingly, in the context of quantum transport~\cite{Politzer, Beenaker1993, Beenaker1993B, Chen, bohigas1, sommers, vivo2008distribution, Khor, kanzieper2, bohigas2, kedar, Texier13, Grabsch1, Grabsch4, Grabsch3} using the random matrix theory approach the conductance ($r(y)= y$~\cite{vivo2008distribution}), Wigner time delay ($r(y)=y$~\cite{Grabsch3}) and shot noise  ($r(y) = y(1-y)$~\cite{kanzieper2}) have also been computed.

In this section, we generalize the results to any arbitrary functions of $r(y)$.  Clearly, the average value of $\langle s\rangle = \langle S_N \rangle/N$, in the large $N$ limit, is given by
\begin{align}
\langle s \rangle= \int_{-l_0}^{l_0} r(y)\,  \rho_{0}(y)\, dy\, ,
\label{mean_s}
\end{align}
with $\rho_{0}(y)$ given explicitly in Eq.~\eqref{rho_uc}. Here, we would like to go beyond the mean $\langle s\rangle$ and compute the variance of $s$ for all $k>1$, which were recently computed for 1dOCP, $k=-1$ (jellium model) in Ref.~\cite{flack2022exact} (see also Ref.~\cite{de2023linear}) and then extended to all long-ranged cases $k<1$ in Ref.~\cite{beenakker2022pair}.

We follow the method used in Ref.~\cite{flack2022exact}. We first compute the full distribution of $s$ in the large $N$ limit. This is done by adding an extra term $\mu_r(s)\, \left( \int_{-\infty}^{\infty} dy ~r(y)\rho_r(y) -s\right)$ in the energy function and then minimizing the energy by the saddle point method. Here $\mu_r(s)$ is the new Lagrange multiplier that enforces the value $s$ of the linear statistics and hence $\mu_r(s)$ depends implicitly on $s$. The subscript `{\it r}' represents the fact that the density and the corresponding chemical potential should depend on the choice of the function $r(y)$. Consequently, the new saddle point density $\rho_r^*(y)$ satisfies the saddle point condition
\begin{align}
\frac{y^2}{2}+ \mu_r^*(s)~r(y) + J \zeta(k)(k+1)~ \big{(}\rho_r^*(y)\big{)}^{k}= \mu_k(s)
\label{dens_mod.1}
\end{align}
where $\mu_k^*(s)$ is the $s$-dependent Lagrange multiplier that enforces the normalization. For the sake of brevity, we omit the $s$ dependence of $\mu_r^*$ and $\mu_k^*$. Thus, the modified density is given by
\begin{align}
\rho_r^*(y)= A_k\, \left(\mu_k^*- \frac{y^2}{2}- \mu_r^*~r(y)\right)^{\frac{1}{k}}\, .
\label{dens_new}
\end{align}
Consequently, the edges of the support, $-l_1(s)$ and $l_2(s)$, where the density vanishes, are determined by the two real roots of
\begin{equation}
\frac{l^2}{2}+ \mu_r^*\, r(l)- {\mu}_k^*=0\, .
\label{support_new}
\end{equation}
The two Lagrange multipliers $\mu_r^*$ and $\mu_k^*$ are then fixed by the two conditions
\begin{align}
 \int_{-l_1(s)}^{l_2(s)}dy~\rho_{r}^*(y)= 1,~\quad~
\int_{-l_1(s)}^{l_2(s)}dy~r(y)\, \rho_{r}^*(y)= s. \label{r_cons}
\end{align}
Clearly, when $s\to \langle s\rangle$, we have $\mu_r^*\to 0$ and ${\mu}_k^*(s) \to \mu_k^*(\langle s\rangle)=\mu_0 = l_0^2/2$ and the density $\rho_r^*(y) \to \rho_0(y)$. We expect the distribution $\mathtt{P}(S_N = s N, N)$ in the large $N$ limit
to have a large deviation form
\begin{align}
\mathtt{P}(S_N = s N, N) \asymp \exp[- N^{1+2\alpha_k}\, \Lambda(s)]\, \, , 
\label{ldv.1}
\end{align}
with the large deviation function given by 
\begin{align}\label{ldflinear}
    \Lambda(s) = G_r[\rho_r^*(y)]-G_0,
\end{align}
where $G_0 $ is given in Eq.~\eqref{derv:unconstrained-action} and the action $G_r[\rho_r^*(y)] = \mathcal{E}_k[\rho^*_r(y)]$ with $\mathcal{E}_k[\rho^*_r(y)]$ given in Eq.~\eqref{E-func-k>1}. The $s$ dependence of $\rho_r^*(y)$ is implicit and comes from the second constraint in Eq.~\eqref{r_cons}.

To compute the explicit expression for the LDF $\Lambda(s)$ [in Eq.~\eqref{ldflinear}], we need to specify the function $r(y)$. However, we can compute an approximate expression for $\Lambda(s)$ for a general function $r(y)$, when $s$ is around its mean value $\langle s\rangle$ [Eq.~\eqref{mean_s}]. When we expand $\Lambda(s)$ [Eq.~\eqref{ldflinear}] for $s = \langle s\rangle +\kappa$ with small $\kappa$, we find that 
\begin{align}
\Lambda(\langle s\rangle +\kappa) \approx \frac{\kappa^2}{2\sigma_r^2}   ~\text{with} ~~
\sigma_r^2= \frac{\mathscr{I}_2\mathscr{I}_0-\mathscr{I}_1^2}{\mathscr{I}_0},
\label{sdv_lin-stat}
\end{align}
where the constants are
\begin{align}\label{I0}
    \mathscr{I}_0 = 2 \frac{A_k}{k} \int_{-l_0}^{l_0} dy~&\left(l_0^2-y^2\right)^{\frac{1}{k}-1},~~ \mathscr{I}_1 = 2 \frac{A_k}{k} \int_{-l_0}^{l_0} dy~r(y)\left(l_0^2-y^2\right)^{\frac{1}{k}-1}\\
    \label{I2}\mathscr{I}_2 &= 2 \frac{A_k}{k} \int_{-l_0}^{l_0} dy~r(y)^2\left(l_0^2-y^2\right)^{\frac{1}{k}-1}.
\end{align}
The quadratic behaviour of the LDF in Eq.~\eqref{sdv_lin-stat} suggests that the typical fluctuations of $S_N$ around its mean value follows a Gaussian distribution given by [see Appendix~\ref{appendlinear}]
\begin{align}\label{lineargauss}
    \mathtt{P}(S_N = (\langle s\rangle+\kappa)N,N)  \asymp  \exp\Bigg(-\frac{N^2\kappa^2}{2~{\rm Var}_S}\Bigg).
\end{align}
where the variance is given by
\begin{align}
    {\rm Var}_S =\frac{N^{\nu_k} \sigma_r^2}{\beta},~~\text{with}~~\nu_k = \frac{2-k}{k+2}.\label{varlinear}
\end{align}
Note that the $N$ dependence of the variance of the linear statistics is universal for any function $r(y)$. As mentioned previously, by construction, the linear statistic captures the behaviour of number and index distribution. Using $r(y) = \Theta(y+w)\Theta(w-y)$ in Eq.~\eqref{sdv_lin-stat}, one can reproduce the variance of the `number' problem as given in Eq.~\eqref{derv:number-var-m}. Interestingly, unlike the short-range case, for the long-range case of the Dyson's log-gas ($k\to 0$) and the 1dOCP ($k=-1$), the behaviour of the linear statistics for the smooth and non-smooth function $r(y)$ differ~\cite{marino14, marino2016number, Flack2022, flack2022exact, Chen}.

\section{Conclusions}
\label{sec6}
In summary, this study provides a detailed analysis of FCS of a confined short-range Riesz gas ($k > 1$) in equilibrium at temperatures $T \sim O(1)$ (where the entropy may be neglected). We focused on the number and the index distribution, which characterize the fluctuations of the number of particles $\mathcal{N}(W, N)$ and $\mathcal{I}(W, N)$, respectively, in two distinct domains, namely  $ [-W,W]$ and $ (-\infty,W]$. We found that the variance of the number of particles in a given domain scales with the system size as $\sim N^{\nu_k}$ with $\nu_k = (2-k)/(k+2)$.  Our study is a major step forward in generalizing results of the Dyson's log-gas and the 1dOCP to broader class of interacting particles {\it i.e.},  Riesz gas systems with $k >1$. We also found that the distribution of the typical fluctuations of both quantities  $\mathcal{N}(W, N)$ and $\mathcal{I}(W, N)$ around their mean values are Gaussian. These results are obtained by computing the large deviation function (LDF) associated with the distribution of these quantities.

We have employed a field theory method similar to the Coulomb gas method to compute the LDFs for two quantities $\mathcal{N}(W, N)$ and $\mathcal{I}(W, N)$. The method involves determining the saddle point density profiles conditioned on a given fraction of particles inside the specified domain. We found that for both cases (`number' and `index'), the saddle point density profiles possess discontinuities at the location of the boundary of the specified domain and exhibit three different kinds of profiles as either $c$ or $w$ is changed, such that it crosses the transition lines indicated in Fig.~\ref{fig:phase}. These three types of configurations display interesting features -- such as discontinuities and emergence of void regions. Our analytical results for the density profiles are in perfect agreement with numerical computations.

These density profiles are then utilized to calculate the LDFs $\Phi(c,w)$ and $\Psi(c,w)$ analytically for the number and index distributions, respectively. The density profiles determine the values of the LDFs in the respective parameter ranges. In particular, one finds that there exists an interesting regime of the parameter ($c\leq \bar{c}(w)$ for fixed $w$ or $w>\bar{w}(c)$ for fixed $c$) in which the saddle point density profile contains a hole (devoid of particles) at the place of the discontinuity. The LDF corresponding to such density profiles undergoes a discontinuous change in the third-order derivative leading to a third-order phase transition. This transition is similar to the third-order transition observed in random matrix theory~\cite{majumdar2014top} and the 1dOCP~\cite{Flack2021}. Apart from exploring the non-analytic properties of the LDFs we have also discussed its various asymptotic forms of the LDF which allowed us to study well-known problems like hole formation probability or complete confinement probability. Additionally, the index problem provided a natural setting for studying the physical properties like the thermodynamic pressure and bulk modulus. 

Our analysis can be easily adapted to other traps of the form $U(x) = \frac{|x|^{\delta}}{\delta}$. The results obtained for these generic traps closely resemble those obtained for the harmonic trap with $\delta=2$. Specifically, it is observed that the fluctuations of the number of particles in the domain $[-W, W]$ or $(-\infty, W]$ are Gaussian and the variance again scales with $N$ as $\sim N^{\nu_k}$. However, now the exponent gets interestingly modified to $\nu_k = 1-\alpha_k\delta = (\delta+k-k\delta)/(k+\delta)$. Moreover, we found that for any $\delta>0$ the non-analytic properties of the LDF remain the same, still displaying the third-order phase transition.

Our investigation raises several interesting questions that can be addressed in the future. One immediate question would be to ask how the results on FCS get modified for the long-range interacting case of the Riesz gas {\it i.e.}, for $k<1$. Another interesting direction would be to investigate FCS for the gas at high temperatures where the entropy and the energy are comparable. This takes place for $T \sim O(N^{2\alpha_k})$ where $\alpha_k = 1/(k+2)$ for $-2<k<1$ and $\alpha_k=k/(k+2)$ for $k>1$~\cite{agarwal2019harmonically}. 

\section{Acknowledgements}
M. K. would like to acknowledge the generous support provided by several funding agencies, including Project 6004-1 of the Indo-French Centre for the Promotion of Advanced Research (IFCPAR), Ramanujan Fellowship (SB/S2/RJN114/2016), SERB Early Career Research Award (ECR/2018/002085), and SERB Matrics Grant (MTR/2019/001101) from the Science and Engineering Research Board (SERB), Department of Science and Technology, Government of India. 
 A. K. would like to acknowledge the support of DST, Government of India Grant under Project No. ECR/2017/000634 and the MATRICS grant MTR/2021/000350 from the SERB, DST, Government of India. M. K., and A. K. acknowledge the Department of Atomic Energy, Government of India, for their support under Project No. RTI4001. S. N. M. and G. S. acknowledge support from ANR Grant No. ANR-23-CE30-0020-01 EDIPS. They also thank P. Le Doussal for useful discussion on related topics. D. M. thanks the support of the Center of Scientific Excellence at the Weizmann Institute of Science.

\appendix
\section{Non-analytic properties of LDF $\Phi(c,w)$}\label{append:asymptotic}

In this appendix, we investigate the non-analytic properties of the LDFs $\Phi(c,w)$ as given in Eqs.~(\ref{phisimc}) and \eqref{phisimw}. We study the series expansions of the LDF $\Phi(c,w)$ around $c = \bar{c}(w)$ for a fixed value of $w$, where $\bar{c}(w)= \left(w/l_0\right)^{\frac{1}{\alpha_k}}$ and  around $w = \bar{w}(c)$ for a fixed value of $c$, where $\bar{w}(c)= c^{\alpha_k}l_0$.  For this analysis, we take advantage of the separable nature of the LDF {\it i.e.} $ \Phi(c,w) = \mathcal{E}_{\rm in}+\mathcal{E}_{\rm out}-G_0$ for the number distribution problem [see Eq.~\eqref{derv:number-ldf1}]. The energy $\mathcal{E}_{\rm in}$ and $\mathcal{E}_{\rm out}$ for the particles inside and outside the box $[-w,w]$ is given by
\begin{align}\label{appendix:energyin}
    \mathcal{E}_{\rm in} &= \begin{cases}G_0 \times c^{2\alpha_k+1} &~\text{for}~c\leq\bar{c}(w)\\
    G_0 \times c^{2\alpha_k+1}\mathscr{J}\left(\dfrac{w}{\bar{l}}\right) &~\text{for}~c>\bar{c}(w)
    \end{cases}~~ \text{and},\\
    \mathcal{E}_{\rm out} &=  G_0 \times (1-c)^{2\alpha_k+1}\mathscr{H}\Big(\frac{w}{l}\Big), \label{appendix:energyout}
\end{align}
where the constant $G_0$ is given in Eq.~\eqref{derv:unconstrained-action} and, the functions $\mathscr{J}(h)$ and $\mathscr{H}(h)$ are given in Eq.~\eqref{derv:number-j} and Eq.~\eqref{derv:number-h}, respectively. Here the length scales $l$ and $\bar{l}$ are given in Eqs.~\eqref{derv:number-suppq} and~\eqref{derv:number-suppm}.

\subsection{$\Phi(c,w)$ for $c\sim\bar{c}(w)$ for a fixed $w$}
\label{phivsc}

We first present the series expansion of the $\mathcal{E}_{\rm in}$ and $\mathcal{E}_{\rm out}$ separately as follows: \\

\noindent
\textit{Energy within the box}~$(\mathcal{E}_{\rm in})$: When the fraction of particles $c$ within the box $[-w, w]$ is slightly greater than the $\bar{c}(w)$, i.e., $c = (1 + \epsilon)\bar{c}(w)$ with $0<\epsilon \ll 1$, we can find the free energy of particles inside the box by expanding the expression in Eq.~\eqref{appendix:energyin} in terms of $\epsilon$. This expansion yields the following series:
\begin{align}
\mathcal{E}_{\rm in} &= G_0 ~\big(\bar{c}(w)\big)^{1+ 2\alpha_k}\Big[1+ \mathbb{A}_{\rm in}~\epsilon + \mathbb{B}_{\rm in}~\epsilon^2 +\mathbb{C}_{\rm in+}~\left(\epsilon\right)^{2+\frac{1}{k}}+o(\epsilon^{2+\frac{1}{k}})\Big],\label{phivsc+ein}~\text{where},\\
\mathbb{A}_{\rm in}& = 1+2\alpha_k,~\quad~
\mathbb{B}_{\rm in} = \frac{(1+2\alpha_k)2\alpha_k}{2},~\quad~
\mathbb{C}_{\rm in+}= \frac{k^2(1+2\alpha_k) \left(2\alpha_k\right)^{2+\frac{1}{k}}}{(k+1)(2k+1)~B\left(\frac{1}{2}, \frac{1}{k}+1\right)}.\label{phivsccin}
\end{align}
To obtain Eq.~\eqref{phivsc+ein} we use the expressions of $\bar{l}$ and $l$ which are functions of $c$ and $w$ and are given in Eqs.~\eqref{derv:number-suppq} and~\eqref{derv:number-suppm}. The notation  small `$o$' specifically $o(\epsilon^{a})$ indicates corrections smaller than $\epsilon^a$ as $\epsilon \to 0$.

Conversely, when the fraction of particles within the box is smaller than $\bar{c}(w)$ and satisfies $c = (1 +\epsilon)\bar{c}(w)$, with $\epsilon<0$ and $|\epsilon| \ll 1$, we find that the energy of the particles inside the box is given by:
\begin{align}\label{phivsc-ein}
\mathcal{E}_{\rm in} &= G_0~\big(\bar{c}(w)\big)^{1+2\alpha_k} \Big[1+ \mathbb{A}_{\rm in}~\epsilon + \mathbb{B}_{\rm in}~\epsilon^2 +\mathbb{C}_{\rm in-}~\epsilon^{3}+O(\epsilon^4)\Big],
\end{align}
where the constants $\mathbb{A}_{\rm in}$ and $ \mathbb{B}_{\rm in}$ are given in Eq.~\eqref{phivsccin}, and  
\begin{align}
\mathbb{C}_{\rm in-} = \frac{(1-4\alpha_k^2)2\alpha_k}{6}.\label{phivsccin-} 
\end{align}

\noindent
\textit{Energy outside the box}~$(\mathcal{E}_{\rm out})$: Similarly, we examine the energy of particles outside the box given in Eq.~\eqref{appendix:energyout}. We find the following expansion in powers of $\epsilon$ 
\begin{align}
\mathcal{E}_{\rm out} &=G_0~\big(1-\bar{c}(w)\big)^{1+2\alpha_k}\Big[\mathcal{H}^{(0)}(\tilde{h})+\mathbb{A}_{\rm out}~\epsilon + \mathbb{B}_{\rm out}~\epsilon^2+\mathbb{C}_{\rm out}~\epsilon^3 + O(\epsilon^4)\Big],\label{phivsc+eout}\\
    \mathbb{A}_{\rm out} &=\tilde{h}~\gamma_{\rm out}~\mathcal{H}^{(1)}(\tilde{h}) -\frac{\bar{c}(w)(1+2\alpha_k)}{1-\bar{c}(w)}, \label{phivscaout}\\
    \mathbb{B}_{\rm out} &=\frac{\big(\bar{c}(w)\big)^2  2\alpha_k(1+2\alpha_k)}{2\big(1-\bar{c}(w)\big)^2}+\frac{\tilde{h}^2\gamma_{\rm out}^2}{2}\mathcal{H}^{(2)}(\tilde{h})-\frac{\bar{c}(w)~h~\gamma_{\rm out}~\mathcal{H}^{(1)}(\tilde{h})(1+2\alpha_k)}{1-\bar{c}(w)},\label{phivscbout}\\
    \mathbb{C}_{\rm out} &=-\frac{\big(\bar{c}(w)\big)^3 2\alpha_k( 4\alpha_k^2-1)}{6\big(1-\bar{c}(w)\big)^3} + \frac{\big(\bar{c}(w)\big)^2~\tilde{h}~ \gamma_{\rm out}~\mathcal{H}^{(1)}(\tilde{h}) 2\alpha_k(1+2\alpha_k)}{2\left(1-\bar{c}(w)\right)^2}+\frac{h^3~\gamma_{\rm out}^3}{6}\mathcal{H}^{(3)}(\tilde{h})\notag\\&-\frac{\bar{c}(w)~\tilde{h}^2~\gamma_{\rm out}^2~\mathcal{H}^{(2)}(\tilde{h})(1+2\alpha_k)}{2\big(1-\bar{c}(w)\big)},\label{phivsccout}
\end{align}
where $\tilde{h} = w/l(\bar{c}(w),w)$ and $\mathcal{H}^{(n)}(\tilde{h})$ is the $n^{\rm th}$ derivative of $\mathcal{H}(\tilde{h})$ given in Eq.~\eqref{derv:number-h}. The constant $\gamma_{\rm out}$ is given by
\begin{align}
\gamma_{\rm out} = \frac{\bar{c}(w)\alpha_k}{1-\bar{c}(w)}\Bigg[\frac{1-I\left(\tilde{h}^{2}, \frac{1}{2}, 1+\frac{1}{k}\right)}{1-I\left(\tilde{h}^{2}, \frac{1}{2}, 1+\frac{1}{k}\right) + 2\alpha_k \frac{\tilde{h}(1-\tilde{h}^{2})^{\frac{1}{k}}}{B\left(\frac{1}{2}, \frac{1}{k}+1\right)}}\Bigg],
\end{align}
where the function $I(g, a, b)$ is given in Eq.~\eqref{I(g,a,b)-def}. 

We can now calculate the LDF $\Phi(\bar{c}(w)(1+\epsilon),w)$ for small $|\epsilon| \ll 1$ by substituting the expressions of the energies from Eqs.~\eqref{phivsc+ein},~\eqref{phivsc-ein} and~\eqref{phivsc+eout} in the expression of the LDF $\Phi(c,w)$  given in Eq.~\eqref{derv:ldf}. This yields:
\begin{align}\label{phic}
    \Phi(\bar{c}(w)(1+\epsilon),w)-\Phi(\bar{c}(w),w) &= 
    \begin{cases}
        \tilde{\mathbb{A}}~\epsilon + \tilde{\mathbb{B}}~\epsilon^2 + \tilde{\mathbb{C}}_{+}~\epsilon^{2+\frac{1}{k}} + o(\epsilon^{2+\frac{1}{k}}),&~\text{for}~\epsilon>0\\
        \tilde{\mathbb{A}}~\epsilon + \tilde{\mathbb{B}}~\epsilon^2+\tilde{\mathbb{C}}_{-}~\epsilon^3 + O(\epsilon^{4}),&\text{for}~\epsilon<0
    \end{cases},
\end{align}
where the constants are given by
\begin{align}
    \tilde{\mathbb{A}} &=G_0\left(\big(\bar{c}(w)\big)^{1+2\alpha_k }\mathbb{A}_{\rm in} + \big(1-\bar{c}(w)\big)^{1+2\alpha_k }\mathbb{A}_{\rm out}\right)\label{atilde},\\
    \tilde{\mathbb{B}} &=G_0\left(\big(\bar{c}(w)\big)^{1+2\alpha_k }\mathbb{B}_{\rm in} + \big(1-\bar{c}(w)\big)^{1+2\alpha_k }\mathbb{B}_{\rm out}\right)\label{btilde},\\
    \tilde{\mathbb{C}}_{+} &=G_0\big(\bar{c}(w)\big)^{1+2\alpha_k}\mathbb{C}_{\rm in+},~\quad~
    \tilde{\mathbb{C}}_{-} =G_0\left(\big(\bar{c}(w)\big)^{1+2\alpha_k }\mathbb{C}_{\rm in-}+\big(1-\bar{c}(w)\big)^{1+2\alpha_k}\mathbb{C}_{\rm out}\right)\label{ctilde-}.
\end{align}
The values of $\mathbb{A}_{\rm in}$, $\mathbb{B}_{\rm in}$, $\mathbb{C}_{\rm in}$, $\mathbb{A}_{\rm out}$, $\mathbb{B}_{\rm out}$ and $\mathbb{C}_{\rm out}$ are provided in Eqs.~\eqref{phivsccin}, ~\eqref{phivscaout}, ~\eqref{phivscbout} and~\eqref{phivsccout}. From Eq.~\eqref{phic}, we find that the third derivative of the LDF $\Phi(c, w)$ w.r.t. $c$ shows a discontinuity at $c = \bar{c}(w)$, which is a signature of a third order phase transition. 

\subsection{$\Phi(c,w)$ for $w\sim\bar{w}(c)$ for a fixed $c$}\label{phivsw}
By following the same procedure as in the previous subsection~\ref{phivsc}, we can expand the LDFs $\Phi(c, (1+\epsilon)\bar{w}(c))$ in powers of $\epsilon$, for small $|\epsilon| \ll 1$ at a fixed $c$. We find:
\begin{align}\label{phiw}
    \Phi(c,\bar{w}(c)(1+\epsilon))-\Phi(c,\bar{w}(c)) &=
    \begin{cases}
        \tilde{\mathbb{D}}~\epsilon + \tilde{\mathbb{E}}~\epsilon^2+ \tilde{\mathbb{F}}_{+}~\epsilon^3 + O(\epsilon^{4}),&~\text{for}~\epsilon>0\\
        \tilde{\mathbb{D}}~\epsilon + \tilde{\mathbb{E}}~\epsilon^2 + \tilde{\mathbb{F}}_{-}~|\epsilon|^{2+\frac{1}{k}} + o(|\epsilon|^{2+\frac{1}{k}}),&\text{for}~\epsilon<0
    \end{cases},
\end{align}
where the constants are given by
\begin{align}
    \tilde{\mathbb{D}} &=G_0\big(1-\bar{c}(w)\big)^{1+2\alpha_k }\tilde{h}~\Gamma_{\rm out}\mathcal{H}^{(1)}(\tilde{h}),~~
    \tilde{\mathbb{E}} =G_0\big(1-\bar{c}(w)\big)^{1+2\alpha_k }\frac{\Gamma_{\rm out}^2\tilde{h}^2}{2}\mathcal{H}^{(2)}(\tilde{h}),\label{etilde}\\
    \tilde{\mathbb{F}}_{+} &=G_0\big(1-\bar{c}(w)\big)^{1+2\alpha_k }\frac{\Gamma_{\rm out}^3\tilde{h}^3}{6}\mathcal{H}^{(3)}(\tilde{h}),~~
    \tilde{\mathbb{F}}_{-} =G_0\frac{\big(\bar{c}(w)\big)^{1+2\alpha_k}k^2(1+2\alpha_k) \left(2\right)^{2+\frac{1}{k}}}{(k+1)(2k+1)~B\left(\frac{1}{2}, \frac{1}{k}+1\right)},\label{ftilde-}
\end{align}
where in this case $\tilde{h} = \bar{w}(c)/l(c, \bar{w}(c))$ and the function $\mathcal{H}^{(n)}(h)$ is the $n^{\rm th}$ derivative of $\mathcal{H}(h)$ given in Eq.~\eqref{derv:number-h}. The constant $\Gamma_{\rm out}$ is given by
\begin{align}
    \Gamma_{\rm out} &= \Bigg[\frac{1-I\left(\tilde{h}^{2}, \frac{1}{2}, 1+\frac{1}{k}\right)}{1-I\left(\tilde{h}^{2}, \frac{1}{2},1+\frac{1}{k}\right) + 2\alpha_k \frac{\tilde{h}(1-\tilde{h}^{2})^{\frac{1}{k}}}{B\left(\frac{1}{2}, \frac{1}{k}+1\right)}}\Bigg].
\end{align}
Here the function $I(g, a, b)$ is given in Eq.~\eqref{I(g,a,b)-def}. From Eq.~\eqref{phiw}, we find that the third derivative of the LDF $\Phi(c, w)$ w.r.t. $w$ shows a discontinuity at $w = \bar{w}(c)$ which is a signature of a third order phase transition.

\section{Pressure and Bulk modulus}
\label{pressureandbulkmodulus}
For the short-range Riesz gas, we find that the index problem provides a natural setup to compute the pressure in the bulk of the gas. Consider the unconstrained Riesz gas in thermal equilibrium characterised by the density profile $\rho_{0}(y)$ from Eq.~\eqref{rho_uc}. The thermodynamic pressure of this gas at a location $W$ can be thought of as the free energy change of the particles to the left of $W$ when they are pushed by moving a wall from $W$ by an infinitesimal amount $\epsilon_N$. Using the separable (additive) nature of free energy in Eq.~\eqref{ldfindex}, one can easily identify the free energy of the left partition which is given by 
\begin{align}\label{derv:free-left}
     \Psi^{(L)}(c^{*}(w), w) = N^{1+2\alpha_k}\frac{G_0}{2}\bigg(\big(2c^{*}(w)\big)^{\frac{3k+2}{k+2}}\mathscr{J}\left(\dfrac{w}{\bar{l}}\right)\bigg),
\end{align}
where $w = W/N^{\alpha_k}$, $\bar{l}$ is given in the Eq.~\eqref{derv:index-suppq} and $c^{*}(w)=\int_{-l_0}^wdy~\rho_0(y)$, represents the fraction of particles below $W$ when the gas is at equilibrium (without any constraint). Also, the function  $\mathscr{J}(h)$ in Eq.~\eqref{derv:free-left} is given in Eq.~\eqref{appendix:ji}.  When the particles on the left are pushed by moving a wall from $W$ to $W-\epsilon_N$ the free energy on the left changes to $\Psi^{(L)}(c^{*}(w), w-\epsilon)$ where $\epsilon = \epsilon_N/N^{\alpha_k}$. Note the fraction of particle on the left of $W-\epsilon_N$ remains the same. The pressure is then obtained by taking a derivative of the free energy of the left partition and can be written as 
\begin{align}\label{derv:index-press}
    \mathbb{P}(W,N) &= N^{1+\alpha_k}\frac{d}{d\epsilon}\Psi^{(L)}(c^{*}(w), w-\epsilon)\Big|_{\epsilon =0}\notag \\ 
    &= N^{\frac{2(k+1)}{k+2}}\mathbf{P}\left(\frac{W}{N^{\alpha_k}}\right), 
\end{align}
where the scaling function $\mathbf{P}(w)$ is given by
\begin{align}\label{derv:index-press1}
    \mathbf{P}(w) = J\zeta(k)k\left(\rho_0(w)\right)^{k+1}.
\end{align}
One can also define a mechanical pressure $\mathbb{P}_{\rm M}(W, N)$ locally at $W$ as the average force exerted by the particles above $W$ on the particles below $W$ and can be expressed as 
\begin{align}
  \mathbb{P}_{\rm M}(W, N) = \Bigg\langle\sum_{i=1}^{s}\sum_{j=s+1}^N \frac{1}{|x_j-x_i|^{k+1}}\Bigg\rangle,
  \label{def:pressure}
\end{align}
where $x_s \leq W <x_{s+1}$ and $x_i$'s are the unscaled positions of the particles. We remark that the definition of the local mechanical pressure does not involve the external potential explicitly, but only implicitly through the average over the equilibrium measure in Eq.~\eqref{def:pressure}. 

\begin{figure}
    \centering
    \includegraphics[scale=0.7]{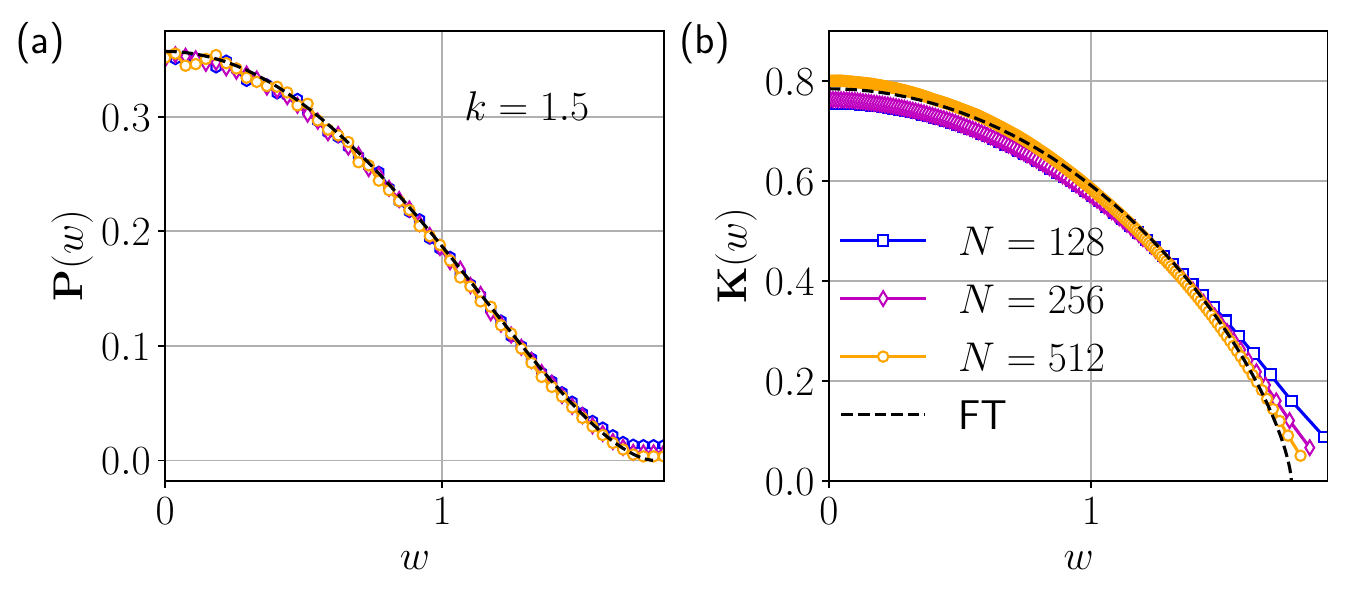}
    \caption{In (a) we plot the scaling function of the thermodynamic pressure [Eq.~\eqref{derv:index-press1}], $\mathbf{P}(w)$, as depicted by the dashed line and the scaled average mechanical pressure [Eq.~\eqref{def:pressure}], $\mathbf{P}_{\rm M}(w N^{\alpha_k}, N)/N^{\frac{2(k+1)}{k+2}}$, as indicated by symbols. The average in Eq.~\eqref{def:pressure} is computed using the MC simulations. In plot (b), the scaling function of the bulk modulus [Eq.~\eqref{scalingcompres}], $\mathbf{K}(w)$, as depicted by the dashed line is compared with $\mathbb{K}_m/N$ [Eq.~\eqref{def:K}] where $m = c^*(w) N$ and is shown by symbols. In Eq.~\eqref{def:K}, the variance of the position of the $m^{\rm th}$ particle is computed using the MC simulations. We use $k=1.5$, $T=1$ and $J=1$ for performing the MC simulations of the unconstrained gas consisting of $N=128, 256$ and $512$ particles and the averages are computed using $10^6$ samples.}
    \label{fig:press-compress}
\end{figure}

We observe that in our model the mechanical pressure $\mathbb{P}_{\rm M}(W, N)$ [see Eq.~\eqref{def:pressure}] and the thermodynamic pressure $\mathbb{P}(W, N)$ [see Eq.~\eqref{derv:index-press}] yield the same result as shown in Fig.~\ref{fig:press-compress}a. Since the gas is confined to a harmonic trap the pressure is not uniform as expected. It is maximum at the centre of the trap and decreases as we go further from the centre of the trap and becomes zero at the edge of the support of the scaled density profile at $y = l_0$ because the value of the density decreases to zero.\\

\noindent
\textit{Bulk modulus:} As usually done in statistical mechanics, to define the bulk modulus we here consider the change of the mean position $\langle x_m\rangle$ of a particle inside the bulk (say the $m^{\rm th}$) due to an external force $F$ applied only on that particle. The bulk modulus is defined as 
\begin{align}
\frac{1}{\mathbb{K}_m} = -\frac{1}{\beta}\frac{\partial \langle x_m \rangle}{\partial F} \Big{|}_{F=0} =  \frac{1}{\beta^2}\partial_F^2 \ln Z(\beta,F)|_{F=0},
\end{align}
where $Z(\beta,F)$ is the partition function of the system in the presence of an external force $F$ on the $m^{\rm th}$ particle {\it i.e.} with the energy function $\tilde{E}_F(\{x_i\}) = \tilde{E}_k(\{x_i\})+F x_m$ that appears in the Gibbs-Boltzmann distribution. A straightforward calculation shows the following fluctuation-response relation
\begin{align}
\frac{1}{\mathbb{K}_m} = \langle x_m^2 \rangle - \langle x_m \rangle^2= N^{2 \alpha_k} (\langle y_m^2 \rangle - \langle y_m \rangle^2),
\label{def:K}
\end{align}
where we have used the scaled position $y_m=x_m/N^{\alpha_k}$. 

To obtain the bulk modulus  at position $W$, denoted by $\mathbb{K}(W, N)$, we need to compute the variance of the position $y_m$ of the $m^{\rm th}$ particle such that $m=c^*(w)N$ with $w=W/N^{\alpha_k}$ and $c^*(w) = \int_{-l_0}^wdy~\rho_{0}(y)$. In other words 
\begin{align}
 \mathbb{K}(W, N) &=  \mathbb{K}_{m=c^*(w)N}~\text{with}~w=\frac{W}{N^{\alpha_k}}. 
 \label{eq:K-K_m}
\end{align}
To  proceed, we first  note that 
\begin{align}\label{probycn1}
    {\rm Prob}.[y_{m} \le w] = \mathcal{P}\left(\mathcal{I} = m, N\right),
\end{align}
where from Section~\ref{indist} we have 
\begin{align}\label{probycn}
    \mathcal{P}\left(\mathcal{I} = m, N\right)  \asymp \exp\left(-\beta N^{1+2\alpha_k}\Psi\left(\frac{m}{N}, w\right)\right).
\end{align}
From this probability distribution, it is straightforward to see that
\begin{align}
\langle y_m^2 \rangle - \langle y_m \rangle^2 =\Bigg(\beta N^{1+2\alpha_k} \frac{d^2}{d\epsilon^2} \Psi(c^*(w), w-\epsilon)\Big|_{\epsilon = 0} \Bigg)^{-1}. 
\label{varsigma_m}
\end{align}
Inserting Eq.~\eqref{varsigma_m} in Eq.~\eqref{def:K}, we get 
\begin{align}\label{derv:index-compress}
    \mathbb{K}(W, N) &=N \mathbf{K}\left(\frac{W}{N^{\alpha_k}}\right),~~\text{where},
\end{align}
\begin{align}\label{scalingcompres}
\mathbf{K}(w)=\frac{k}{(k+2)}\left[\frac{l_0^2\left(\rho_{0}(w)\right)^2}{\bigg(c^{*}(w)-\frac{k}{k+2}w\rho_{0}(w)\bigg)\bigg(1-\Big(c^{*}(w)-\frac{k}{k+2}w\rho_{0}(w)\Big)\bigg)} \right].
\end{align}
In Fig.~\ref{fig:press-compress}b, $\mathbf{K}(w)$ given in Eq.~\eqref{scalingcompres} is plotted and compared with $\mathbb{K}_m$ , obtained from MC simulations. We observe a good agreement as stated in  Eq.~\eqref{eq:K-K_m} which improves as $N$ is increased. In this figure, we observe that the bulk modulus monotonically decreases starting from a finite value at the centre of the trap and approaches zero at the edge of support of the scaled density profile $\rho_0(y)$ at $l_0$. Near the edge of the scaled density profile, $l_0$, the bulk modulus $\mathbf{K}(w \to l_0)\sim (l_0-w)^{\frac{1}{k}}$ and its derivative exhibits a discontinuity. This reflects a third-order phase transition, interpreting the bulk modulus as an order parameter.

\section{Variance of linear statistics}\label{appendlinear}
In this section, we study the variance of linear statistics $S_N = \sum_{i=1}^N r(y_i)$ [Eq.~\eqref{r_def}] for arbitrary function $r(y)$.  Note that $r(y) = \Theta(w-y)\Theta(w+y)$ and $r(y) = \Theta(w-y)$ correspond the number problem and index problem, respectively. To compute the variance we adapt the method of the field theory described in Ref.~\cite{santra2022}. We are computing the typical fluctuations of $S_N$ around its mean  $\langle S_N \rangle  = \langle s\rangle N$ where $\langle s\rangle$ is given in Eq.~\eqref{mean_s}. We express $S_N$ as $S_N = N (\langle s\rangle +\kappa)$ where $N \kappa$ is the fluctuation. To compute the variance, we consider the small deviations, $\Delta \rho_r \ll 1$, of the density profile from the unconstrained density profile as
\begin{align}\label{densitylinearappe}
    \rho_r(y) = \rho_0(y)+\Delta\rho_r(y).
\end{align}
Note that the deviation in density profile satisfies the constraints in Eq.~\eqref{r_cons}, which becomes
\begin{align}\label{norm.1appe}
    \int_{-l_0}^{l_0}dy~\Delta\rho_r(y) = 0,~~\quad~~
    \int_{-l_0}^{l_0}dy~r(y)\Delta\rho_r(y) = \kappa,
\end{align}
where the limits of the integrals have been approximated to leading order.
Using the Eq.~\eqref{densitylinearappe} in the saddle point equation Eq.~\eqref{dens_mod.1} and assuming the contribution due to the higher order terms $O((\Delta\rho_r^*)^3)$ are negligible, gives
\begin{align}\label{deltadenslinearappe}
    \Delta\rho_r^*(y) = \frac{\Delta\mu_k^* + r(y)\Delta\mu_r^*}{J\zeta(k)(k+1)(k)}\left(\rho_0(y)\right)^{1-k},
\end{align}
where `$*$' represents the saddle point value. Here $\Delta \mu_k^* = \mu_k^*-\mu_0$ and $\Delta\mu_r^* = \mu_r^*$. Inserting the expression of the perturbed density from  Eq.~\eqref{deltadenslinearappe} in the constraints in Eq.~\eqref{norm.1appe} we get
\begin{align}\label{constraint_app1}
    \Delta\mu_k^* \mathscr{I}_0+\Delta\mu_r^* \mathscr{I}_1 = 0,~~\quad~~
    \Delta\mu_k^* \mathscr{I}_1+\Delta\mu_r^* \mathscr{I}_2 = \kappa,
\end{align}
with the constants $\mathscr{I}_0, \mathscr{I}_1, \mathscr{I}_2$ given explicitly in Eq.~\eqref{I0} and~\eqref{I2} which we recall to be
\begin{align}\label{I0_app}
    \mathscr{I}_0 = 2 \frac{A_k}{k} \int_{-l_0}^{l_0} dy&~\left(l_0^2-y^2\right)^{\frac{1}{k}-1},~~
\mathscr{I}_1 = 2 \frac{A_k}{k} \int_{-l_0}^{l_0} dy~r(y)\left(l_0^2-y^2\right)^{\frac{1}{k}-1},\\
    \label{I2_app}\mathscr{I}_2 &= 2 \frac{A_k}{k} \int_{-l_0}^{l_0} dy~r(y)^2\left(l_0^2-y^2\right)^{\frac{1}{k}-1}.
\end{align}
Solving for $\Delta \mu_k^*$ and $\Delta \mu_r^*$ in Eq.~\eqref{constraint_app1} one finds
\begin{align}\label{deltamuklinearappe}
    \mu_k^* = \mu_0 -\frac{\kappa \mathscr{I}_1}{\mathscr{I}_2\mathscr{I}_0-\mathscr{I}_1^2},~~\quad~~
    \mu_r^* = \frac{\kappa \mathscr{I}_0}{\mathscr{I}_2\mathscr{I}_0-\mathscr{I}_1^2}.
\end{align}
Inserting the expression for the perturbed density $\rho_r^*(y) = \rho_{0}(y) + \Delta \rho_r(y)$ with $\Delta \rho_r(y)$ from Eq.~\eqref{deltadenslinearappe} and the perturbed chemical potentials from Eq.~\eqref{deltamuklinearappe}, in the expression of the LDF given in Eq.~\eqref{ldflinear}, we find that $\Lambda(s=\langle s\rangle +\kappa)$ (upto quadratic order in $\kappa$) is given by
\begin{align}\label{ldflinearquad}
    \Lambda(s = \langle s\rangle +\kappa) = \frac{\kappa^2}{2 \sigma_r^2}~\text{where}~\sigma_r^2 = \frac{\mathscr{I}_2\mathscr{I}_0-\mathscr{I}_1^2}{\mathscr{I}_0}.
\end{align}
Here the constants $\mathscr{I}_0$, $\mathscr{I}_1$ and $\mathscr{I}_2$ are given in Eqs.~\eqref{I0_app} and~\eqref{I2_app}. By substituting this LDF [Eq.~\eqref{ldflinearquad}] in large deviation form given in Eq.~\eqref{ldv.1}, we find that the typical fluctuations of $S_N$ are Gaussian distributed [see Eq.~\eqref{lineargauss}] with the variance given by [Eq.~\eqref{varlinear}]
\begin{align}\label{varlinearapp}
    {\rm Var}_r = \frac{N^{1-2\alpha_k}\sigma_r^2}{\beta}.
\end{align}
Here $\sigma_r^2$ is given in Eq.~\eqref{ldflinearquad} and it depends on the function $r(y)$. By choosing the function $ r(y) = \Theta(y+w)\Theta(w-y)$ in Eq.~\eqref{varlinearapp}, we recover the variance for the number problem as given in Eq.~\eqref{derv:number-var}. Similarly, for the index case, when we choose the function $r(y) = \Theta(w-y)$, the variance is given by
\begin{align}
	{\rm Var}(\mathcal{I}) = \frac{N^{\nu_k}}{\beta~l_0^{2}~\alpha_k}~\mathcal{U}\left(\frac{W}{N^{\alpha_k}l_0}\right),\label{derv:index-vars}
\end{align}
where $l_0$ is given in Eq.~\eqref{A_1k} and the exponent $\nu_k = 1-2\alpha_k = (2-k)/(2+k)$. Here the function $\mathcal{U}(h)$ is given by
\begin{align}\label{U(h)}
    \mathcal{U}(h) = \frac{\left(1-I\left(h^2, \frac{1}{2}, \frac{1}{k}\right)\right)~\left(1+I\left(h^2, \frac{1}{2}, \frac{1}{k}\right)\right)}{4},
\end{align}
where  the function $I(h,a,b)$ is defined in Eq.~\eqref{I(g,a,b)-def}.

\begin{center}
\line(1,0){250}
\end{center}

\section*{References}
\bibliography{ref.bib}

\end{document}